\documentclass[aps,prf,preprint,longbibliography,superscriptaddress]{revtex4-2}
\usepackage{amsmath}
\usepackage{amssymb}
\usepackage{amsfonts}
\usepackage{color}
\usepackage{graphicx,caption,subfig}
\usepackage{bm}

\usepackage[export]{adjustbox}

\newcommand{\Wi}{\mathrm{Wi}}
\renewcommand{\Re}{\mathrm{Re}}

\begin{document}

\title{Narwhals and their blessings: \protect\\exact coherent structures of elastic turbulence in channel flows}

\author{Alexander Morozov}
\email{alexander.morozov@ed.ac.uk}
\affiliation{School of Physics and Astronomy, The University of Edinburgh, James Clerk Maxwell Building, Peter Guthrie Tait Road, Edinburgh, EH9 3FD, United Kingdom}

\author{Martin Lellep}
\affiliation{School of Physics and Astronomy, The University of Edinburgh, James Clerk Maxwell Building, Peter Guthrie Tait Road, Edinburgh, EH9 3FD, United Kingdom}

\author{Damiano Capocci}
\affiliation{School of Physics and Astronomy, The University of Edinburgh, James Clerk Maxwell Building, Peter Guthrie Tait Road, Edinburgh, EH9 3FD, United Kingdom}

\author{Moritz Linkmann}
\affiliation{School of Mathematics and Maxwell Institute for Mathematical Sciences, \\ University of Edinburgh, Edinburgh, EH9 3FD, United Kingdom}

\date{\today}

\begin{abstract}
Solutions of long, flexible polymer molecules are complex fluids that simultaneously exhibit fluid-like and solid-like behaviour. When subjected to external flows, dilute polymer solutions develop elastic turbulence -- a unique chaotic flow state absent in Newtonian fluids such as water. Unlike turbulence in Newtonian fluids, elastic turbulence arises from polymer stretching and alignment in the flow, and can occur even at vanishing inertia. While experimental realisations of elastic turbulence are well documented, its underlying mechanism remains poorly understood.

In this paper, we present a perspective on the transition to elastic turbulence in pressure-driven channel flows, drawing on recent computational work from our group. We outline our current understanding of the transition in both two and three spatial dimensions, centred on two key building blocks: (i) \emph{narwhals}, exact coherent states of the flow, and (ii) \emph{blessings}, spatio-temporal intermittent states made up of several localised narwhal solutions.

This contribution is based on a talk given by one of us (A.M.) at the 2024 APS DFD meeting.
\end{abstract}

\maketitle

\section{Introduction}
\label{sec:intro}

Here, we report on our recent progress in understanding the transition to elastic turbulence in pressure-driven channel flows of dilute polymer solutions. We summarise our current understanding of \emph{narwhals}, exact coherent states of such flows, and their \emph{blessings}, a term borrowed from the collective noun for narwhals, which we use to describe a spatially and temporally intermittent state comprising several localised narwhal solutions. The paper combines previously published and new computational results, and is based on a talk given by one of us (A.M.) at the 77th Annual Meeting of the APS Division of Fluid Dynamics, held in Salt Lake City, Utah, in 2024.

Elastic turbulence is a chaotic flow regime observed in non-Newtonian fluids~\cite{Steinberg2021,Datta2022}. It is characterised by a broad range of spatial and temporal fluctuations caused by the fluid’s intrinsic elasticity, which stems from the stretching and reorientation of its microstructural constituents under flow~\cite{Perkins1995}. This elasticity is quantified by the Weissenberg number, $\Wi = \lambda \dot\gamma$, which compares a characteristic deformation rate, $\dot\gamma$, with the fluid's relaxation time $\lambda$~\cite{Bird1987}. For $\Wi>1$, the flow velocity gradients are sufficiently strong to stretch the fluid’s microstructure well beyond its equilibrium configuration, with the principal stretch direction aligned with the flow. This results in strongly anisotropic material properties and a mismatch between the stresses along the flow and gradient directions. Absent in Newtonian fluids, this mismatch, known as the (first) normal stress difference, is a defining characteristic of viscoelasticity and is the driving force behind elastic turbulence~\cite{Bird1987}. 

In the more familiar case of hydrodynamic turbulence, the transition is governed by the Reynolds number, $\Re$, which compares inertial effects to viscous stresses~\cite{SchmidHenningson2000}. In Newtonian fluids such as water, turbulence emerges only for sufficiently large $\Re\gg 1$. In contrast, elastic turbulence is driven by large anisotropic stresses and does not require inertia; it can arise at arbitrarily small $\Re$, earning it the byname of \emph{turbulence without inertia} ~\cite{Larson2000}, as long as $\Wi>1$. Here, we focus on the purely elastic regime, characterised by large $\Wi$ and $\Re<1$. 

Although elastic turbulence was observed experimentally in a variety of complex fluids, ranging from liquid crystals~\cite{Kang2006,Wychowaniec2021,Datta2022} to worm-like micelles~\cite{Fardin2010,Datta2022}, dilute polymer solutions are the main experimental system used to study it~\cite{Steinberg2021}. Chaotic flows of polymer solutions have been (re-)discovered several times in the past 100 years (see Dubief et al.~\cite{Dubief2023} for a nice summary of the historical literature). In the modern context, elastic turbulence was discovered by Groisman and Steinberg~\cite{Groisman2000}, who performed first controlled experiments with properly characterised fluids, and helped to establish elastic turbulence as a novel field of rheology and soft condensed matter. 

For some time, it was believed that the emergence of elastic turbulence requires curved streamlines in the base flow. Such flows were known to exhibit a linear instability driven by the elastic tension in the streamlines~\cite{Shaqfeh1996}, or the \emph{hoop stresses}, that are responsible for the rod-climbing effect~\cite{Bird1987} (or for the bread dough climbing on the kneading hook). Their mechanism is summarised by the Pakdel-McKinley criterion~\cite{McKinley1996,Datta2022} that expresses the critical Weissenberg number in terms of the ratio of the flow curvature and the distance travelled by a polymer molecule in one relaxation time. Such \emph{shear} instabilities (as opposed to the \emph{extensional} ones that are the main focus of this paper), were shown to give rise to complicated secondary flows that, in turn, became unstable for slightly larger $\Wi$, leading eventually to elastic turbulence. This transition scenario is well-verified~\cite{Groisman2001,Schiamberg2006} although a direct, sub-critical transition to elastic turbulence in such flows was also reported~\cite{Groisman2004}.

In the case of parallel shear flows, however, the Pakdel-McKinley criterion predicts the absence of a linear instability and this conclusion seemed to be supported by the early linear stability analyses~\cite{Ho1977,Wilson1999} (see also Castillo S\'{a}nchez \emph{et al.}~\cite{Sanchez2022} for a review). In the absence of the streamline curvature and the associated linear instability, viscoelastic parallel flows were believed to be stable until Professor Wim van Saarloos (Leiden University) and Professor Daniel Bonn (University of Amsterdam) suggested that elastic turbulence in such flows could originate through a \emph{bifurcation from infinity}~\cite{Rosenblat1979}, i.e. be triggered by a finite-amplitude perturbation. This proposal, put forward in detail in~\cite{Morozov2007}, resulted in a series of systematic experimental studies broadly confirming the presence of an instability in such flows~\cite{Bertola2003,Bonn2011,Pan2013,Qin2017,Qin2019,Steinberg2022}, although the recent results of the Steinberg group have opened a possibility of a linear transition~\cite{Shnapp2022,Li2023,Li2024,Li2024a,Li2025}. 

Until recently, computational understanding of viscoelastic parallel shear flows remained very limited. The first breakthrough was achieved by Berti \emph{et al.}~\cite{Berti2008,Berti2010} in the context of model two-dimensional Kolmogorov flow. There, Berti and colleagues reported the first coherent state in such flows that they simply referred to as elastic waves. Their discovery was not appreciated at the time and it is only now that their work is beginning to receive the recognition it deserves. (Remarkably, a few years later, Zhang et al.~\cite{Zhang2013a} performed three-dimensional simulations of visco-elastic Kolmogorov flow at low values of $\Re\sim O(1)$ but the relatively low numerical resolution and the lack of detailed stress visualisations likely prevented them from identifying the three-dimensional analogues of these structures that are the main subject of our paper.) The next major development was due to Shankar and colleagues who demonstrated that channel and pipe flows of model polymeric fluids do, in fact, exhibit a linear instability, despite the earlier reports mentioned above~\cite{Ho1977,Wilson1999}. In contrast to their Newtonian counterparts~\cite{SchmidHenningson2000}, that are driven by a wall-based mechanism, these \emph{centremode} instabilities where first reported in elasto-inertial flows (moderate $\Wi$ and $\Re$)~\cite{Garg2018,Chaudhary2019,Khalid2021a,Chaudhary2021} and then tracked down to the purely elastic regime~\cite{Khalid2021,Buza2022,Buza2022a}. In purely elastic pressure-driven channel flow, the resulting linear instability is confined to a part of the parameter space with very high $\Wi$ and very small polymer concentration, while keeping $\Re\ll 1$. Although the direct observation of a linear instability in this regime is challenging, its existence is crucial in understanding the transition. This was convincingly demonstrated by Page \emph{et al.}~\cite{Page2020}, who showed that the non-linear state emerging from the centremode instability in the elasto-inertial regime is sub-critical and can persist even for the linearly stable conditions. The corresponding non-linear states, that the authors named \emph{arrowheads}, are now understood to be identical to the elastic waves of Berti \emph{et al.}~\cite{Berti2008,Berti2010}. Importantly, this independent discovery linked these states to the centremode linear instability at moderate $\Wi$ and $\Re$. The consecutive studies by the same group have further investigated how these elastic waves or arrowheads can lead to chaotic behaviour in two-dimensional channel flows~\cite{Dubief2022,Beneitez2024,Beneitez2024a}.

In this paper, we review the recent work that took place in our group in parallel with the development discussed above. Motivated by our early proposal that elastic turbulence is organised around linearly unstable exact coherent states and guided by the works of Berti \emph{et al.}~\cite{Berti2008,Berti2010} and Page \emph{et al.}~\cite{Page2020}, we first reported the purely elastic channel flow analogues of the elastic waves/arrowhead structures \cite{Morozov2022} (Section \ref{sec:2D}). Throughout this paper, we follow the suggestion of Professor Becca Thomases (Smith College) and Professor James Hanna (University of Nevada) and refer to these coherent states as \emph{narwhals}, reflecting the close analogy between the stress distribution associated with these structures and the body shape of the animal; additionally, this choice provides a convenient language to describe the mechanism of how they are sustained, as we show below. Importantly, we showed that the two-dimensional narwhal states are linearly unstable, when embedded in three spatial dimensions \cite{Lellep2023}, and in Section \ref{subsec:lininstab} we argue that elastic turbulence could only be studied in three-dimensional simulations. The report of the first of such simulations~\cite{Lellep2024} in Section \ref{pnassection} constitutes the final part of our paper.

\subsection{Governing equations}

Unlike their Newtonian counterparts, complex fluids, in general, and polymer solutions, in particular, are not described by a unique set of equations of motion. Instead, each particular system has to be treated on an \emph{ad hoc} basis where one strives to choose a model incorporating the main physics of the system in question. For dilute polymer solutions, the key ingredients are the normal stress difference, responsible for elastic instabilities and turbulence, and shear-thinning, describing how the fluid's viscosity and relaxation time decrease with the deformation rate. Here, we select the simplified Phan-Thien Tanner (sPTT) constitutive model~\cite{PhanThien1977} that incorporates all these ingredients and is well-established in the rheological literature. Other models representing the same physics are available in the literature~\cite{Larson2013,Alves2021}; reviews of the general physics of constitutive equations and recipes of how to choose them can be found elsewhere~\cite{Tanner1985,Bird1987,Larson1999,Morozov2015}.

\begin{figure}[t!]
\centering
\includegraphics[width=0.35\columnwidth]{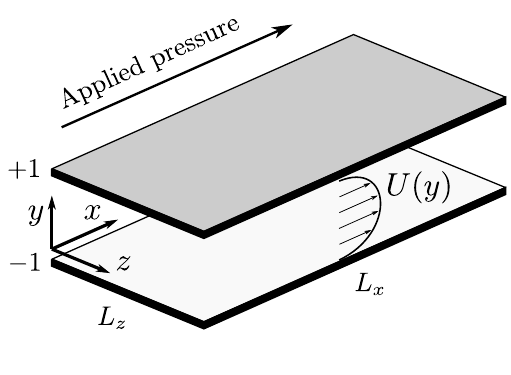}
\caption{Flow geometry studied in this work. $x$, $y$, and $z$ are Cartesian coordinates aligned with the streamwise, wall-normal, and spanwise directions, respectively. The flow is driven in the $x$-direction by a constant, externally applied pressure gradient. In the two-dimensional case, the flow is assumed to be translationally invariant along the spanwise direction.}
\label{geometry}
\end{figure}

In this paper, we study a model sPTT fluid confined within a straight, three-dimensional channel formed by the gap between two parallel, infinite plates. The geometry is conveniently described using a Cartesian coordinate system $(x,y,z)$, aligned with the streamwise, wall-normal, and spanwise directions, respectively; see Fig.\ref{geometry}. A constant external pressure gradient is imposed along the streamwise direction to drive the flow. The dimensionless equations of motion are given by
\begin{align}
\label{eq:ptt} 
& \partial_t {\bm c}
+ {\bm v}\cdot\nabla{\bm c} - \left(\nabla {\bm v}\right)^T\cdot{\bm c} - {\bm c}\cdot\left(\nabla {\bm v}\right) = \kappa \nabla^2 {\bm c} 
- \frac{{\bm c}-\mathbb{I}}{\Wi}\Bigg[ 1 + \epsilon\, \mathrm{Tr}\left({\bm c}-\mathbb{I}\right)\Bigg],  \\
& \label{eq:ns} 
\partial_t  {\bm v} + {\bm v}\cdot\nabla{\bm v}  =
 -\nabla p + \frac{\beta}{\Re} \nabla^2{\bm v} + \frac{1-\beta}{\Re\,\Wi}\nabla\cdot{\bm c} + \frac{2}{\Re}\hat{\bf x}, \\
&\label{eq:incomp} 
\nabla\cdot {\bm v} = 0.
\end{align}
Here, $\bm c$ is the polymer conformation tensor, $\bm v$ is the fluid velocity, $p$ is the pressure, and $\hat{\bf x}$ is a unit vector in the streamwise direction. The superscript $(\cdot)^T$ denotes the transpose, while $\mathbb{I}$ denotes the unit matrix. In this paper, we consider both two- and three-dimensional flows.

The parameter space of this model is spanned by the values of $\Wi$, $\Re$, the shear-thinning parameter $\epsilon$, reduced polymer diffusivity $\kappa$, and the viscosity ratio, $\beta = \mu_s/(\mu_s+\mu_p)$, where $\mu_s$ and $\mu_p$ are the solvent and polymeric contributions to the total viscosity, respectively; $\beta$ can also be viewed as a proxy for the polymer concentration. The way these equations are rendered dimensionless is discussed in Appendix \ref{app:units}.

The simulation domain is chosen to be periodic along the streamwise and spanwise directions with $L_x$ and $L_z$ being the corresponding domain lengths. At the walls, we impose no-slip boundary conditions for the velocity, ${\bm v}(x,y=\pm 1,z,t)=0$,  while the boundary conditions for the conformation tensor are obtained by requiring that $\bm c$ at the walls is equal to the corresponding value obtained by solving Eq.~\eqref{eq:ptt} with $\kappa=0$ \cite{Thomas2006}. 

The equations of motion, Eqs.~\eqref{eq:ptt}-\eqref{eq:incomp}, are solved numerically using an MPI-parallel fully dealiased pseudo-spectral code developed within the Dedalus framework \cite{Burns2020}. Time-discretisation is performed using either a 4th-order semi-implicit BDF scheme \cite{Wang2008} or a four-stage, third-order implicit-explicit Runge-Kutta method \cite{Ascher1997} with the timestep $dt$. The velocity, conformation tensor, and pressure fields are represented through a spectral decomposition based on Fourier-Chebyshev-Fourier modes in the streamwise, wall-normal, and spanwise directions, respectively. The spectral resolution is set by specifying the number of modes, $(N_x,N_y,N_z)$, used in each direction. The values of the parameters and the resolution used in each simulation are discussed in the relevant sections; additional numerical details can be found in the corresponding references. 

\section{Two-dimensional coherent structures}
\label{sec:2D}

As discussed in the Introduction, our first step in understanding the pathway to elastic turbulence in parallel shear flows was spurred by the work of Berti \emph{et al.}~\cite{Berti2008,Berti2010} and Page \emph{et al.}~\cite{Page2020}, together with the discovery of novel purely elastic linear instability by Khalid \emph{et al.}~\cite{Khalid2021}. As in those studies, we first consider two-dimensional pressure-driven flows before studying their three-dimensional counterpart in the next Section. Unless explicitly stated, all runs in this Section employed $L_x=10$, $\Re=10^{-2}$, $\epsilon = 10^{-3}$ and $\kappa = 5\cdot 10^{-5}$, while varying $\beta$ and $\Wi$. We have verified that the base flow is linearly stable for all parameters considered here. 

\subsection{How to find a narwhal}
\label{subsec:2dreport}
Most of the results presented here were originally reported in \cite{Morozov2022}.

\begin{figure}[ht!]
\centering
\includegraphics[width=0.8\columnwidth]{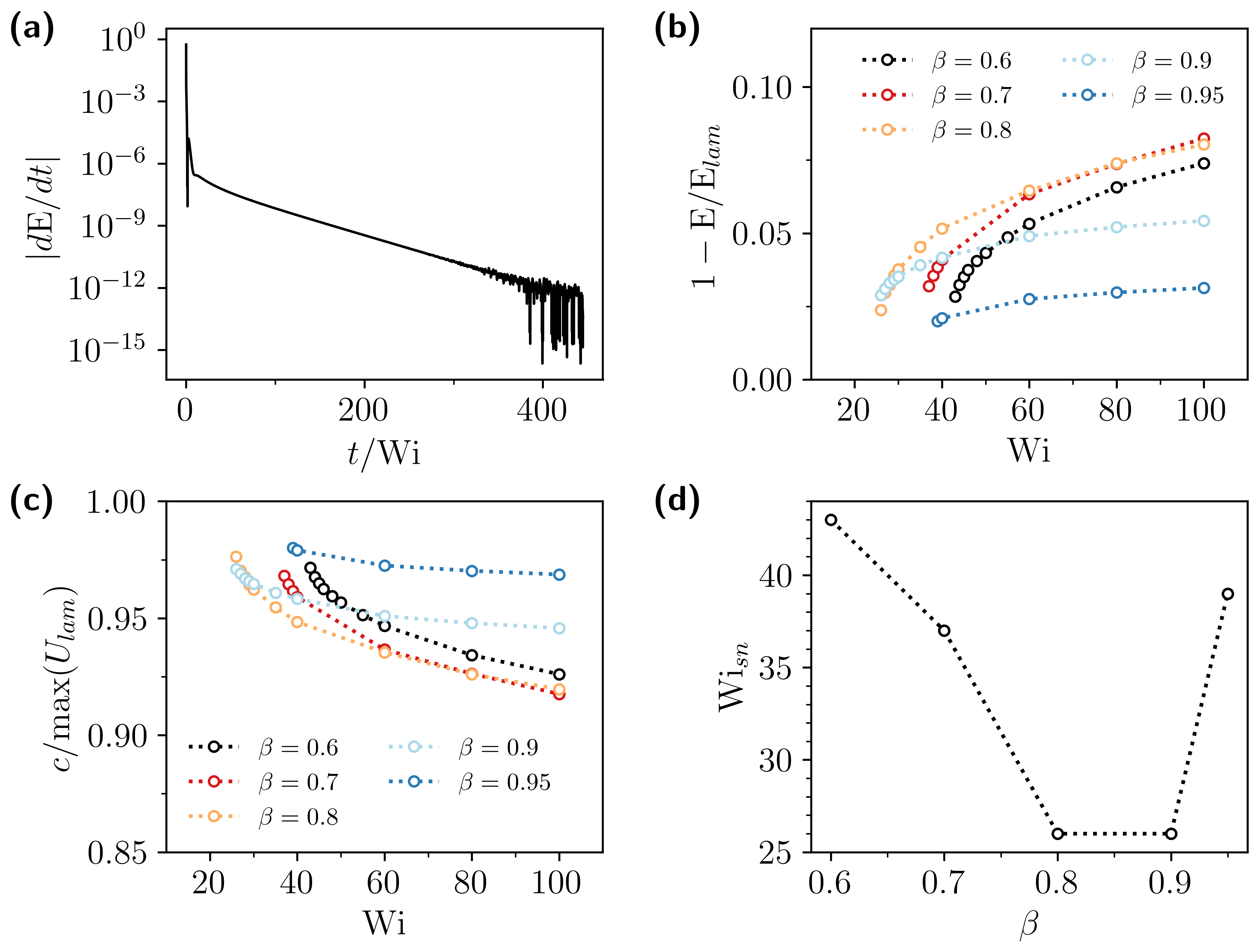}
\caption{
Two-dimensional travelling-wave solutions. (a) Evolution of the time derivative of the kinetic energy, $dE/dt$, as a function of time for $\beta=0.8$ and $\Wi=26$. The steady-state is reached when $\vert dE/dt \vert$ decreases sufficiently to become comparable with the machine precision. 
(b) The reduced kinetic energy of the travelling wave solutions representing the upper branch of the corresponding bifurcation from infinity. 
(c) The speed of the travelling wave solutions determined by tracking the position of the maximum of $\mathrm{Tr}\, {\bm c}$ as a function of time. 
(d) The saddle-node values $\Wi_{sn}$ determined as the lowest value of the Weissenberg number at which the solution can be sustained.
Panels (b)-(d) are replotted from the data originally presented in Fig.1 of Morozov \cite{Morozov2022}.}
\label{fig_PRLcombi}
\end{figure}

To study the non-linear dynamics of two-dimensional pressure-driven channel flows of polymeric fluids, we performed a series of direct numerical simulations varying $\beta$ and $\Wi$. The non-linear states discussed below can be reached by adding a small but finite amount of Gaussian noise to the conformation tensor. However, such perturbations often led to prohibitively long transients. We found instead that the same states can be triggered far more efficiently by choosing initial conditions with a strong extensional component along the channel centreline. In practice, we superimposed a spatially localised spot in the $c_{xx}$ component of the conformation tensor onto the laminar profile.

\begin{figure}[ht!]
\centering
\includegraphics[width=0.8\columnwidth]{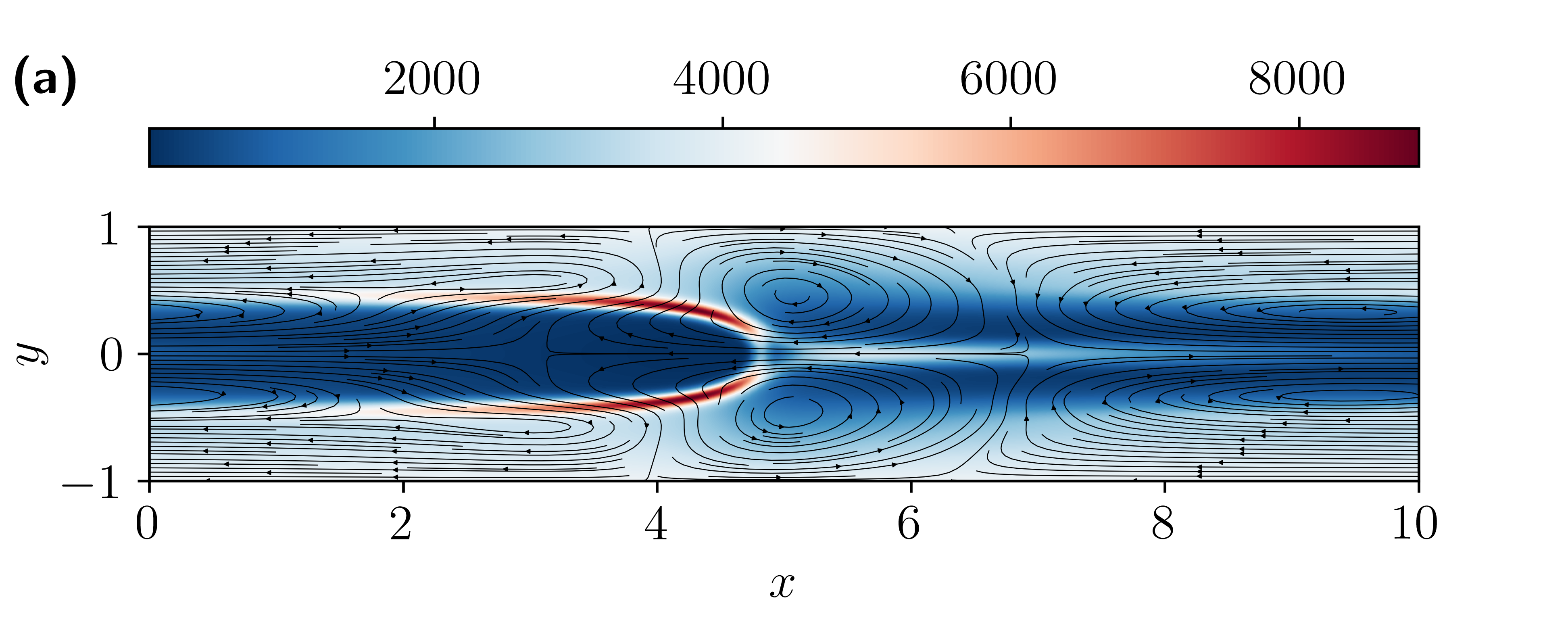}
\includegraphics[width=0.8\columnwidth]{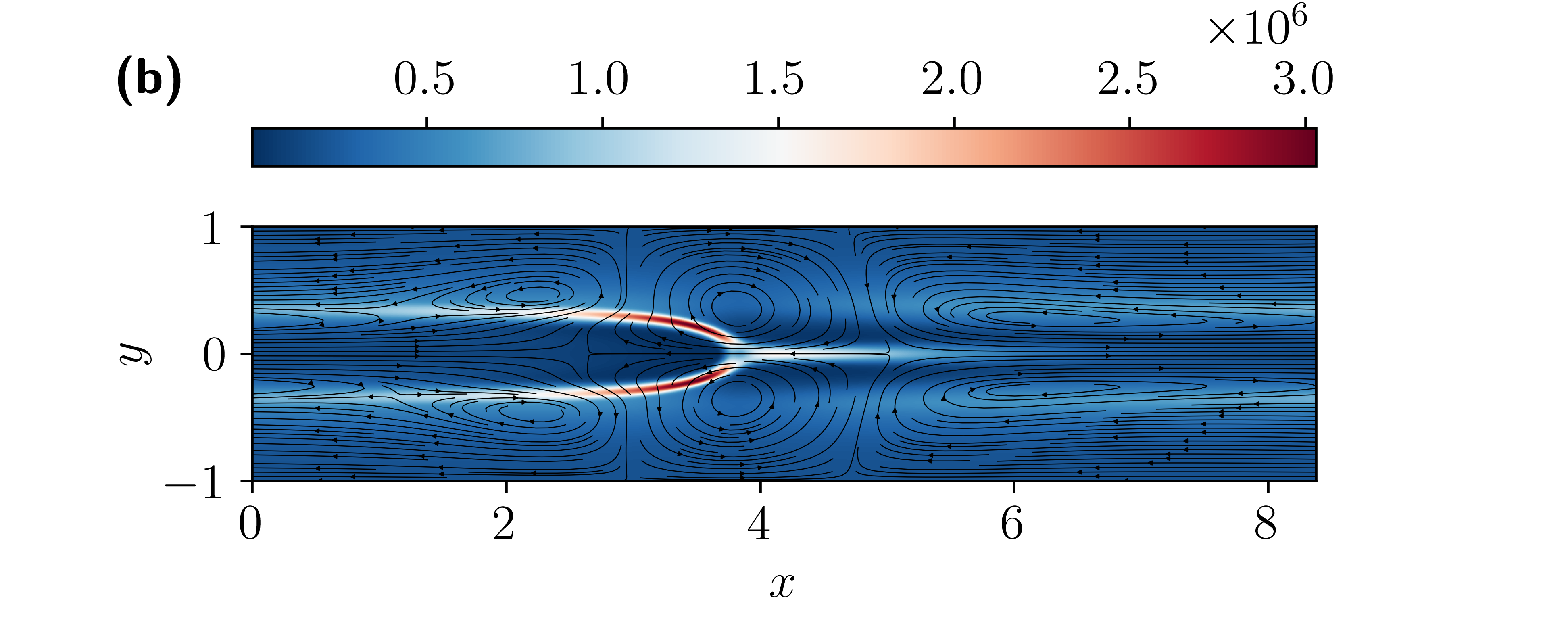}
\caption{Stress and velocity profile of the exact travelling-wave solutions. The polymer stretch (colour) and the flow streamlines (solid lines) for (a) $\beta=0.8$ and $\Wi=100$ and (b) $\beta=0.997$, $\Wi=1700$ and $L_x=8\pi/3$. The mean flow is from left to right. The streamlines represent velocity deviation from the mean streamwise profile. Panels (a) and (b) are replotted from the data originally presented in Fig.1 of Morozov~\cite{Morozov2022} and Fig.1 of Lellep \emph{et al.}~\cite{Lellep2023}, respectively.}
\label{fig_PRLnarwhals}
\end{figure}

In all cases studied here, the time evolution followed a similar trend: after an initial transient, the system converged to a steady state. The nature of this state depends on the Weissenberg number: for sufficiently small $\Wi$ the system relaminarises, whereas above a critical $\Wi$ it settles into a dynamical steady state comprising a flow structure travelling downstream with a constant speed. In Fig.~\ref{fig_PRLcombi}(a) we illustrate this behaviour by plotting the time derivative of the kinetic energy, $dE/dt$, which decays to below machine precision once a dynamical steady state is reached.

Since the laminar flow is linearly stable for the parameters considered, the observed non-linear states arise through a subcritical bifurcation from infinity. We therefore identify the critical $\Wi$ with the saddle-node value $\Wi_{sn}$. In Fig.~\ref{fig_PRLcombi}(b), we construct the bifurcation diagram for this flow by plotting the reduced kinetic energy $1-E/E_{lam}$ versus $\Wi$, where $E$ denotes the kinetic energy of the steady state and $E_{lam}$ its laminar value. 

The non-linear states on the upper branches correspond to travelling-wave solutions: stress and velocity fields that translate downstream at constant speed and are thus steady in a co-moving frame. Their propagation speed, measured by tracking the streamwise position of the maximum polymer stretch $\mathrm{Tr}\,{\bm c}$, is shown in Fig.~\ref{fig_PRLcombi}(c). As in Newtonian flows \cite{Graham2021}, these travelling waves move more slowly than the laminar profile. Finally, Fig.~\ref{fig_PRLcombi}(d) shows $\Wi_{sn}$ as a function of $\beta$, mapping the region of existence of purely elastic travelling-wave states.

In Fig.\ref{fig_PRLnarwhals}(a), we show the stress and velocity profiles associated with the non-linear travelling waves presented above. The distinct feature of this state is the presence of thin, filament-like arrangements of the polymer stretch. Its distinct and unique shape is rather unusual. While Newtonian coherent structures and their self-sustaining mechanism are associated with the presence of boundaries \cite{Graham2021}, the structure reported here is localised around the centreline of the channel. In what follows, we refer to such states as \emph{narwhals}, in view of the strong visual resemblance of the stress profile in Fig.\ref{fig_PRLnarwhals}(a) and the shape of the famous arctic animal \emph{Monodon monoceros}. (One might note the further analogy between the rare sightings of narwhals in the wild and the historical difficulties of finding non-linear solutions in viscoelastic parallel shear flows.)

The narwhal states reported here bear close similarity to the elastic waves found by Berti \emph{et al.}~\cite{Berti2008,Berti2010} in two-dimensional Kolmogorov flow, as well as to the arrowhead structures reported by Page \emph{et al.}~\cite{Page2020} in pressure-driven channel flow at moderate values of $\Re$ and $\Wi$, i.e. in the elasto-inertial regime. The latter were shown to propagate from the centremode linear instability of elasto-inertial flows ~\cite{Garg2018,Chaudhary2019,Khalid2021a,Chaudhary2021}. That instability has since been tracked to the purely elastic regime, $\Re\ll 1$, ~\cite{Khalid2021,Buza2022,Buza2022a}, where it persists only for very large values of $\Wi$ and very small values of $1-\beta$. In Fig.\ref{fig_PRLnarwhals}(b), we report a narwhal state that we calculated in that regime. The strong resemblance between that solution and the narwhals that we found at more moderate, physically relevant values of $\beta$ suggests that they, together with the arrowhead states of Page \emph{et al.}~\cite{Page2020}, belong to a single family of solutions emerging from the centremode linear instability.

\subsection{How to sustain a narwhal}
\label{subsec:explanation}

The shape of the narwhal states presented above is quite peculiar, and its origin is not immediately evident. Here, we explore a self-sustaining mechanism that can stabilise such structures. We demonstrate that, rather than being governed by the classical elastic instability of curved streamlines~\cite{McKinley1996,Datta2022}, the mechanism is extensional in nature, broadly connected to purely extensional flows near stagnation points. Instead of developing a rigorous mathematical theory, we offer a qualitative, pictorial description of the self-sustaining process. An abridged version of this argument was previously presented in \cite{Morozov2022arxiv}.

\begin{figure}[t!]
\centering
\begin{tabular}{cc}

\begin{tabular}{c}
\includegraphics[width=0.48\columnwidth]{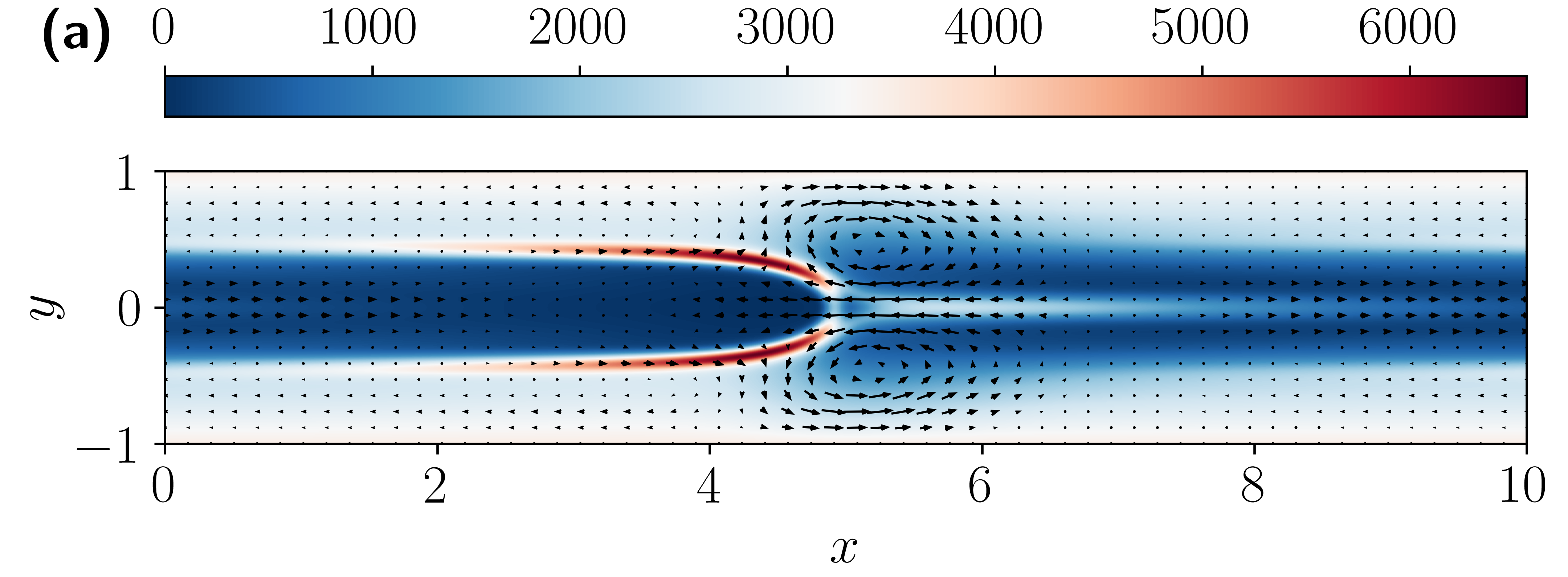}\\
\includegraphics[width=0.48\columnwidth]{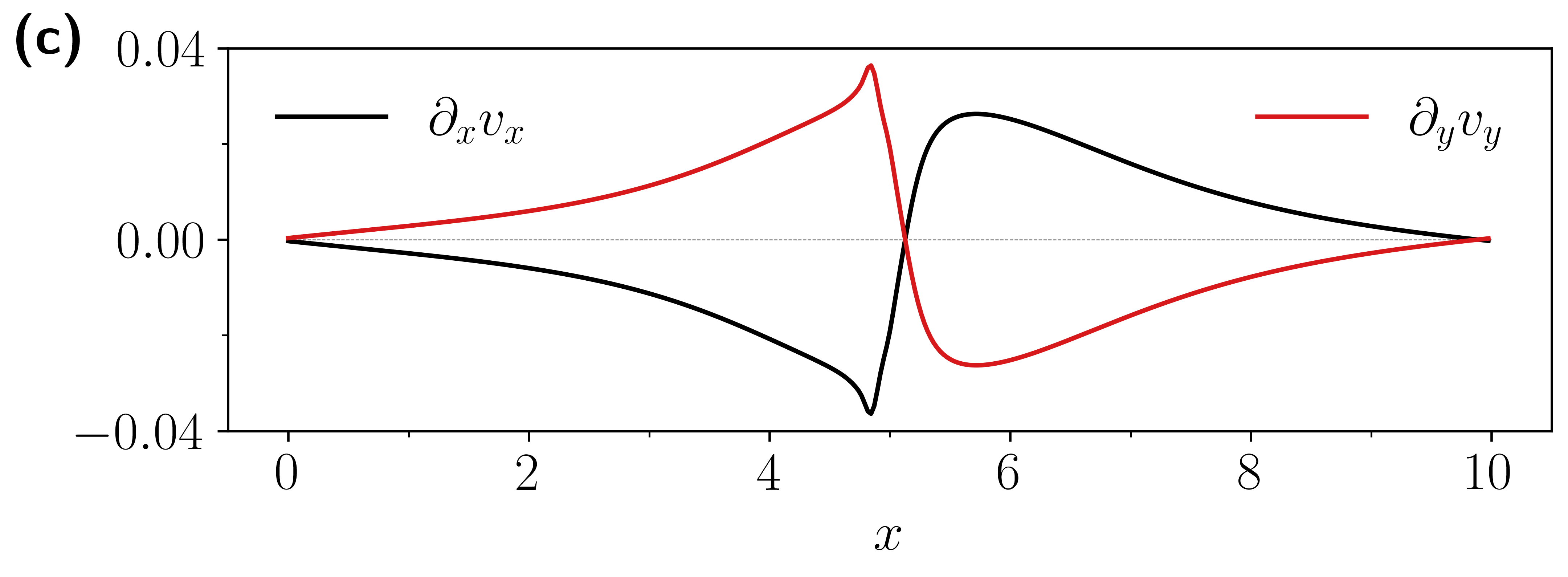} \\
\includegraphics[width=0.30\columnwidth]{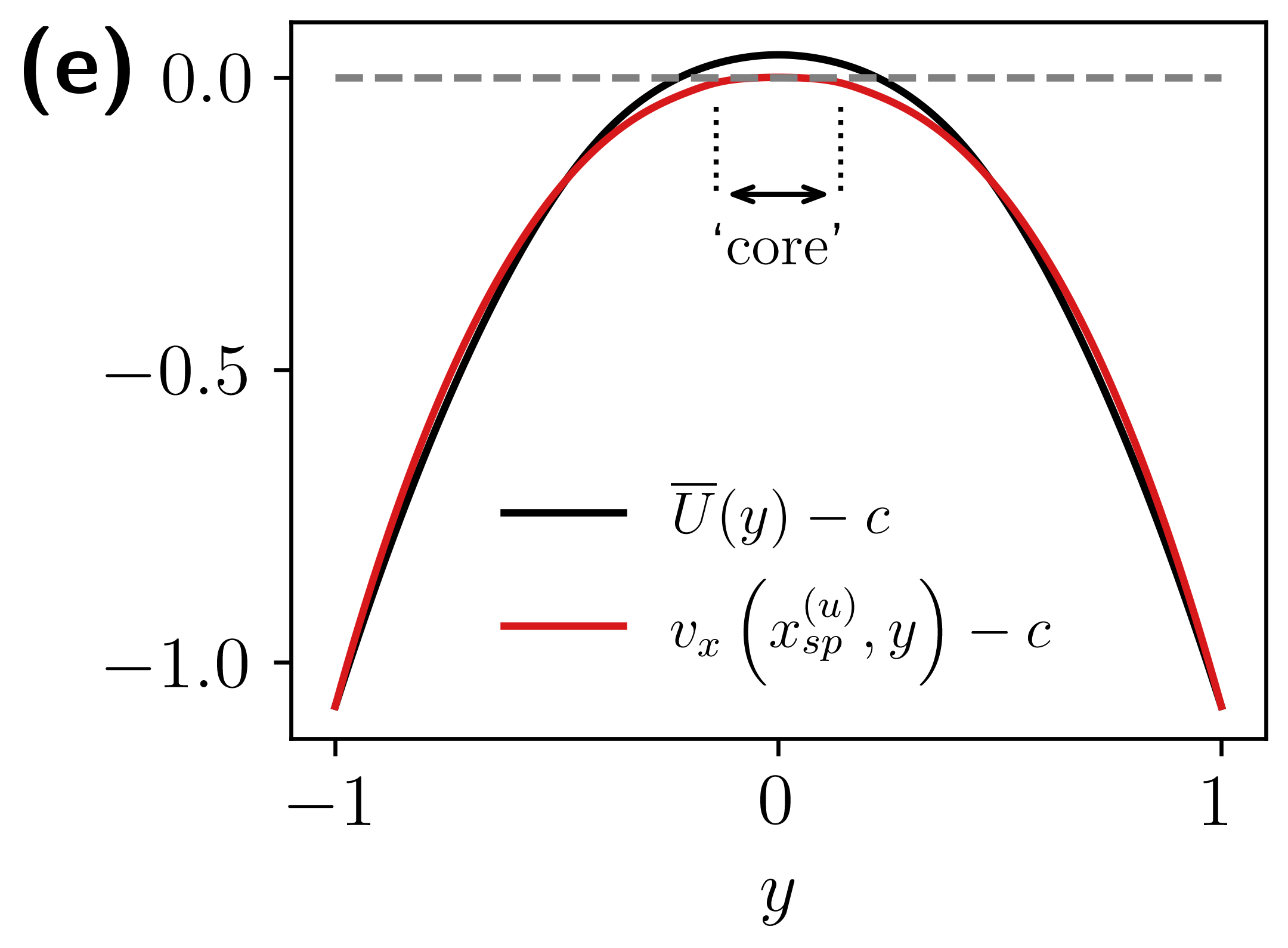}
\end{tabular} & 

\begin{tabular}{c}
\includegraphics[width=0.48\columnwidth]{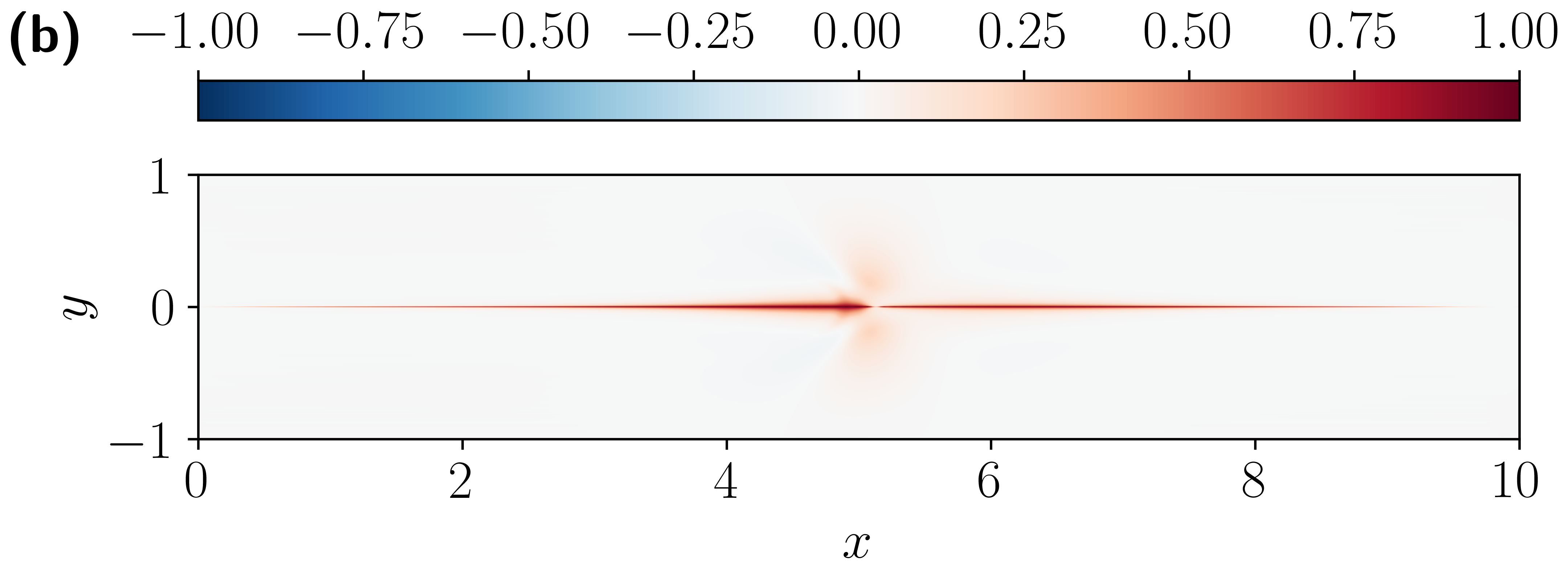}\\
\includegraphics[width=0.48\columnwidth]{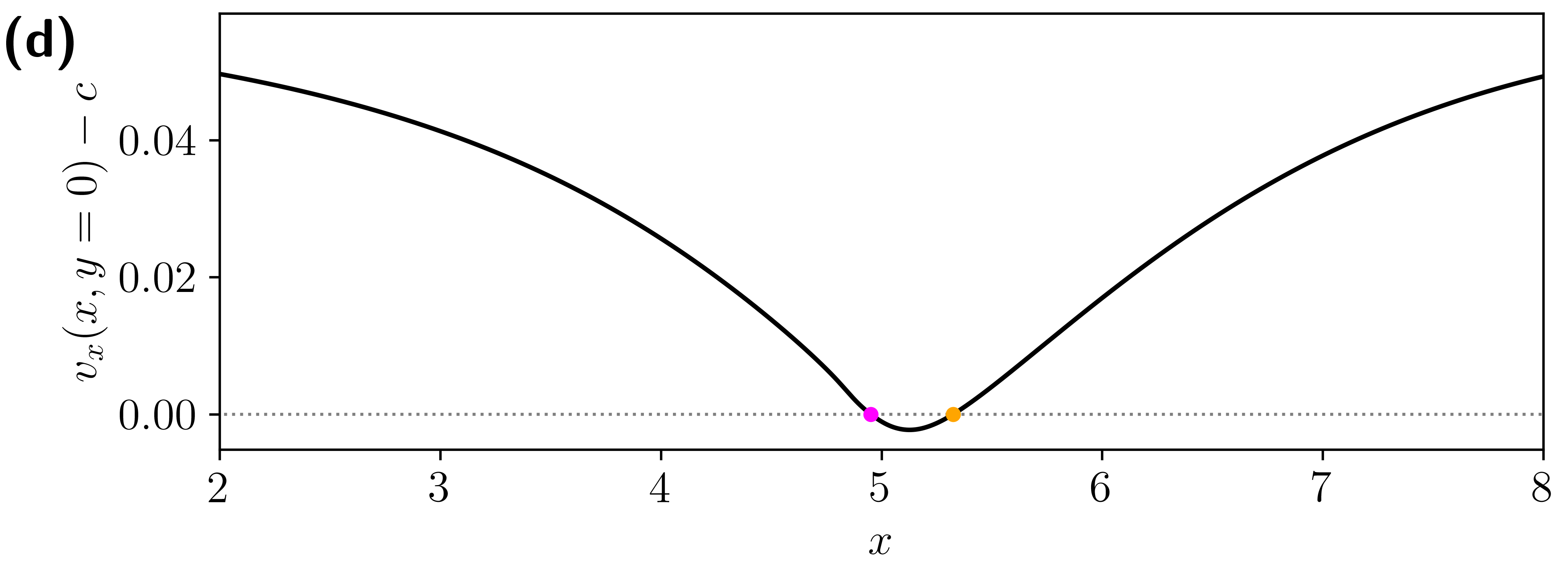}\\
\includegraphics[width=0.55\columnwidth]{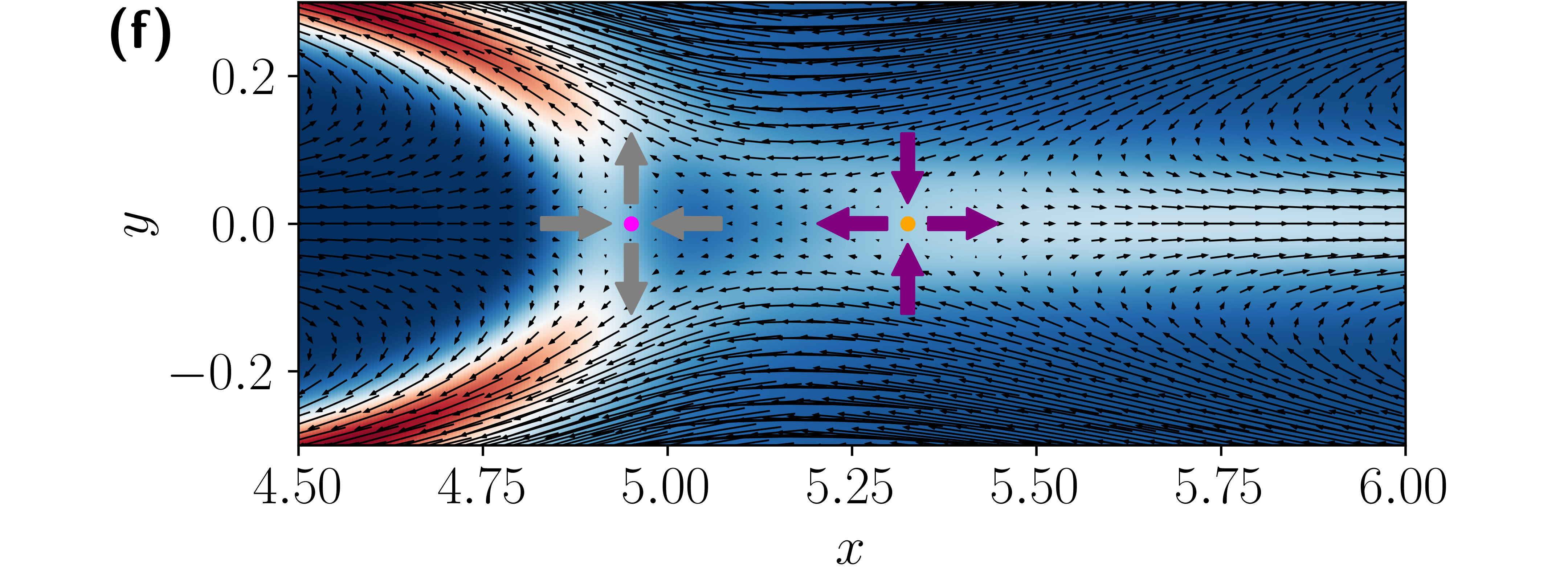}
\end{tabular} 

\end{tabular}
\caption{Characterisation of the narwhal velocity field. This figure is based on the $\beta=0.8$ and $\Wi=80$ data set originally reported in Morozov~\cite{Morozov2022}. (a) Polymer stretch (colour) and the velocity deviation from its mean profile (vectors). (b) Flow type parameter $\chi$. (c) Centreline profiles of the velocity gradient components. Note that $\partial_y v_y = -\partial_x v_x$ due to incompressibility, while $\partial_x v_y = \partial_y v_x = 0$ along the centreline. (d)  The co-moving frame streamwise velocity along the centreline, $v_x(x,y=0)-c$. (e) The co-moving frame mean velocity profile and the local streamwise velocity profile at the position of the upstream stagnation point. (f) Zoom on a part of panel (a) showing the position of the two stagnation points and the local geometry of the flow. Note that the velocity field (vectors) is plotted in the co-moving frame.}
\label{fig_explanation_velocity}
\end{figure}

We start by analysing the velocity field associated with the narwhal state. In Fig.\ref{fig_explanation_velocity}(a), we plot the velocity deviation from its mean profile, ${\bm v}'\left(x,y\right) = {\bm v}\left(x,y\right) - \left(\overline U(y),0\right)$, superimposed onto the spatial profile of $\mathrm{Tr}\,\bm c$, where $\overline U(y)$ is the mean streamwise velocity profile, and $\overline{\left( \dots \right)}$ denotes the average along the $x$-direction. Note, that since the narwhals are travelling-wave solutions, these profiles are time-independent in a co-moving frame (see below).
First, we observe that ${\bm v}'$ is largely confined to a solitary pair of vortices located at the point where the tusk of the narwhal is attached to its body, see Fig.\ref{fig_explanation_velocity}(a). Such structures resemble `diwhirls' previously reported in purely elastic Taylor-Couette flows \cite{Groisman1997,Kumar2000}, although their orientation with respect to the gradient direction is different from the Taylor-Couette case. To understand the nature of this velocity field, we introduce the flow type parameter $\chi$ based on the invariants of the velocity gradient tensor,
\begin{align}
\chi = \frac{\sqrt{\mathbf{D} : \mathbf{D}}   - \sqrt{\mathbf{\Omega} : \mathbf{\Omega}^T}}{\sqrt{\mathbf{D} : \mathbf{D}} + \sqrt{\mathbf{\Omega} : \mathbf{\Omega}^T}},
\end{align}
where $\mathbf{D} = \left( \mathbf{\nabla}\bm{v} + \mathbf{\nabla}\bm{v}^T\right)/2$ and  $\mathbf{\Omega} = \left( \mathbf{\nabla}\bm{v} - \mathbf{\nabla}\bm{v}^T\right)/2$ are the strain and vorticity tensors, respectively.
For model flows, $\chi$ takes well-defined values \cite{Lee2007}: purely extensional flow corresponds to $\chi=1$, shear flow -- to $\chi=0$, and the solid-body rotation -- to $\chi = -1$. In Fig.\ref{fig_explanation_velocity}(b), we plot the flow type parameter $\chi$ calculated for the narwhal state and observe that most of the domain is occupied by simple shear with $\chi=0$, stemming from $\partial_y \overline U(y)$. Since the latter is largely dominated by $\partial_y U^{(lam)}(y)$, this is a feature of the laminar profile. Along the centre line, however, we observe regions of pure extension with $\chi=1$, located in the vicinity of the solitary vortex pair. It is important to stress that the flow-type parameter $\chi$ is insensitive to the local strength of the velocity field, and should be complemented by observations of the local magnitude of $\nabla \bm v$. To this end, in Fig.\ref{fig_explanation_velocity}(c), we plot $\partial_x v_x$ and $\partial_y v_y = -\partial_x v_x$ along the channel centre line, while the other components of the velocity gradient tensor are zero along the centreline.
This allows us to conclude that significant polymer extension is mainly localised along the centreline, around the region where the tusk is attached to the narwhal's body.

For the next step in the argument, we recall that narwhals are travelling-wave solutions, moving downstream with the speed $c$. We now consider a Galilean transformation to a frame moving with the narwhal, $x\to x-c\,t$. As can be seen from Fig.\ref{fig_explanation_velocity}(e), the mean velocity profile in this frame, $\overline{U}(y)-c$, consists of a small region around the centreline that moves downstream, while the bulk of the channel moves in the opposite direction. This behaviour, however, only holds on average, while the local streamwise velocity shows significant variations along the centreline. To corroborate this point, in Fig.\ref{fig_explanation_velocity}(d) we plot the streamwise velocity in the co-moving frame along the centreline, $v_x(x,y=0)-c$, and observe that it vanishes at two points. These points, $x_{sp}^{(u)}$ and $x_{sp}^{(d)}$, denoted by the pink and orange dots in Fig.\ref{fig_explanation_velocity}(d), correspond to the region where the tusk is attached to the narwhal's body, and to the beginning of the tusk, respectively. At these points, the streamwise velocity profile across the gap is nearly zero around the centreline, as illustrated in 
Fig.\ref{fig_explanation_velocity}(e) for $x=x_{sp}^{(u)}$, and we refer to this loosely defined region as the `core'. Taken together, these observations imply that, in the co-moving frame, the velocity field around these two points can be approximated by ${\bm v} \approx \dot\epsilon^{(u)} \left(-(x-x_{sp}^{(u)}),y\right)$ and ${\bm v} \approx \dot\epsilon^{(d)} \left(x-x_{sp}^{(d)},-y\right)$, for $y$ in the core region around the centreline. These profiles correspond to planar extensional flows with $\left(x_{sp}^{(u)},y=0 \right)$ and $\left(x_{sp}^{(d)},y=0 \right)$ defining the upstream (body) and the downstream (tusk) stagnation points, with $\epsilon^{(u)}$ and $\epsilon^{(d)}$ being the corresponding (constant) extension rates. As illustrated in Fig.\ref{fig_explanation_velocity}(f), the flow around the upstream stagnation point contracts along the $x$ and expands along the $y$-direction, while the flow around the downstream stagnation point is doing the opposite.

\begin{figure}[t!]
\centering
\includegraphics[width=0.48\columnwidth]{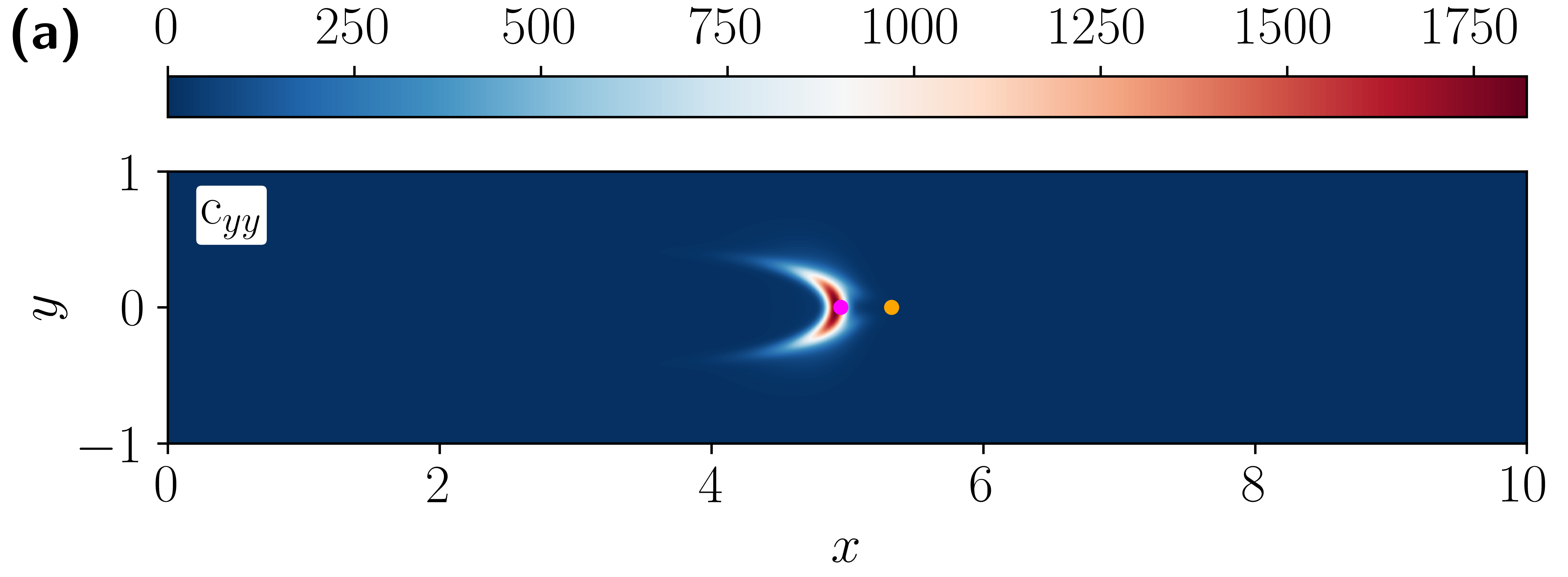}
\includegraphics[width=0.48\columnwidth]{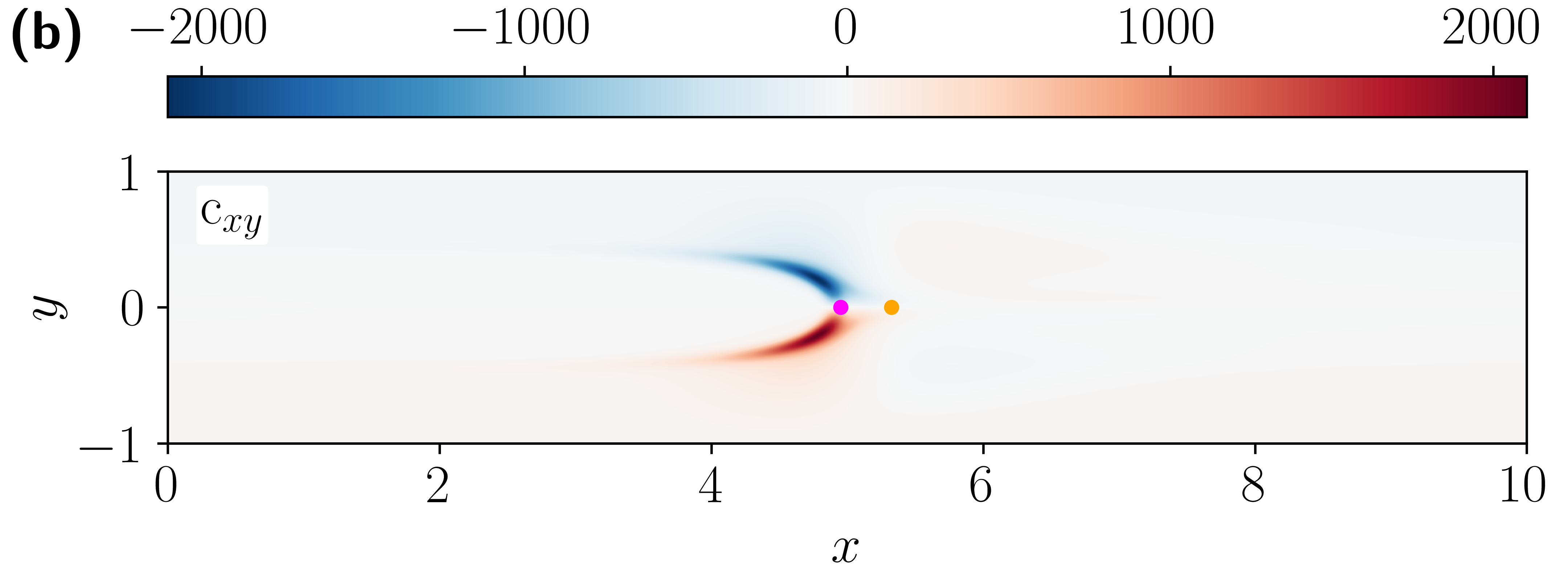}
\includegraphics[width=0.48\columnwidth]{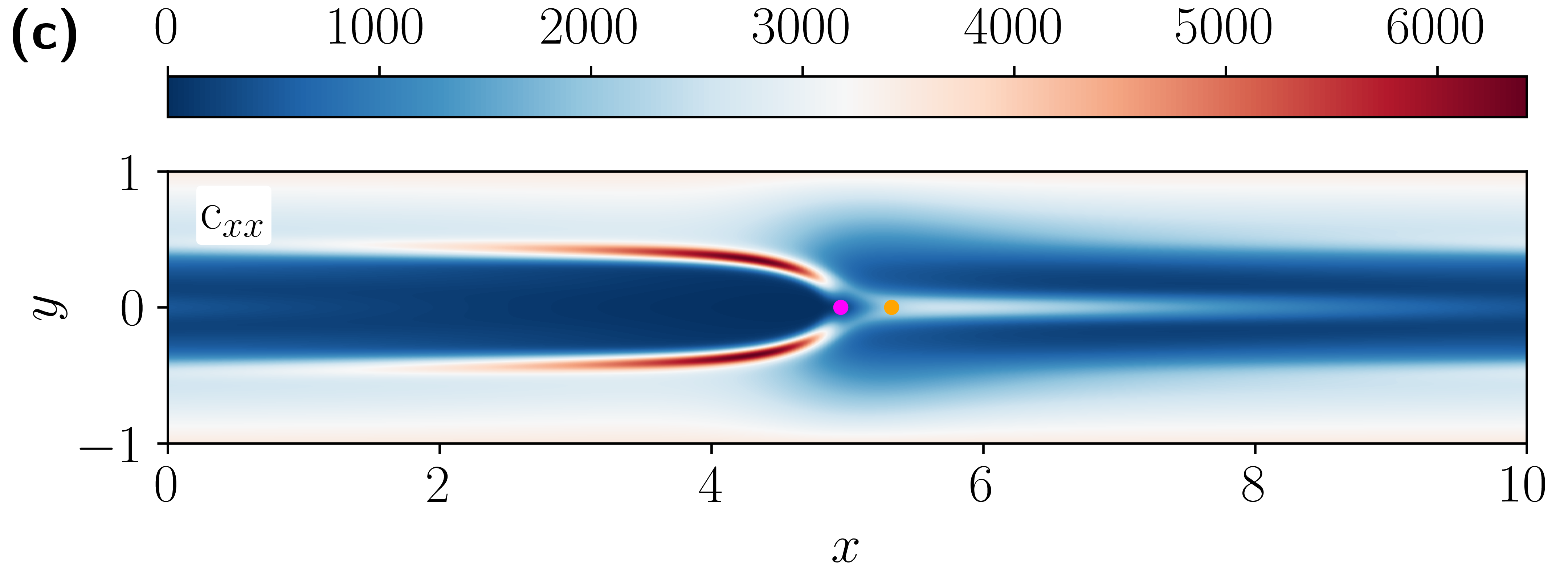}
\caption{Components of the conformation tensor corresponding to the narwhal state presented in Fig.\ref{fig_explanation_velocity}.}
\label{fig_explanation_stress}
\end{figure}

Having established that the narwhal velocity field is controlled by two stagnation points, we now turn to the discussion of the associated stress structure. Previous theoretical studies of planar extensional flows demonstrated the existence of large polymer stresses along the extensional direction of the flow \cite{Rallison1988,Renardy2006,Thomases2006,Becherer2009}. Such stresses are strongly localised along the compression flow direction, revealing themselves as narrow birefringent strands in experiments \cite{Crowley1976,Fuller1980}. When adapted to our coordinate system, this suggests a presence of a significantly large $c_{yy}$ component of the conformation tensor around the upstream stagnation point. This is indeed confirmed in Fig.\ref{fig_explanation_stress} where we plot individual components of $\bm c$. As can be seen from Fig.\ref{fig_explanation_stress}(a), the spatial profile of $c_{yy}$ around the upstream stagnation point indeed exhibits a nearly vertical, filament-like region localised along the $x$-direction; note that the small curvature of the $c_{yy}$ filament for small $y$ is associated with the small, but non-zero curvature of the local streamwise velocity profile in the core region, see Fig.\ref{fig_explanation_velocity}(e). When the $c_{yy}$ filament extends into the bulk beyond the core region, it is being advected by the mean velocity profile that points upstream in the co-moving frame, as discussed above. Since the bulk is dominated by shear generated by the mean profile, see Fig.\ref{fig_explanation_velocity}(b), the extensional mechanism responsible for the production of $c_{yy}$ ceases to be effective outside the core, and $c_{yy}$ decays along the $x$-direction, upstream from the body stagnation point. 

In turn, the presence of a region of large $c_{yy}$, combined with the vorticity, largely dominated by the mean vorticity $\overline U'(y)$, drives the shear component of the conformation tensor through $c_{xy}\sim \Wi \overline{U}' c_{yy}$, where we only indicate the forcing term in the $xy$-component of the constitutive equation. This forcing is only significant in the spatial region where $c_{yy}$ and $\overline{U}'$ are simultaneously non-vanishing; outwith that region, $c_{xy}$ is being advected downstream while decaying. Finally, the same mechanism is responsible for the production of the $xx$-component of the conformation tensor, which is mainly driven by the coupling between the local mean-shear vorticity and $c_{xy}$ through $c_{xx}\sim \Wi \overline{U}' c_{xy}$. The latter, being the final result of the flow-induced alignment of an extended high-stress filament, is oriented almost horizontally. Every step in this process makes polymers stretch and re-orient in the flow, culminating in the $c_{xx}$ component dominating the total polymer stretch throughout the domain. 

Finally, the extensional flow around the downstream stagnation point is responsible for the generation of the filamentous region of large polymer stretch that we identify as the tusk. Its origins lie in the same mechanism as the one responsible for the generation of the vertical arrangement of $c_yy$ around the upstream stagnation point, see Fig. \ref{fig_explanation_stress}(a). The orientation of the outgoing flow around the downstream stagnation point ensures that the tusk is confined in the vertical direction and is oriented along the centreline, with $c_{xx}$ being the dominant component of the conformation tensor, see Fig. \ref{fig_explanation_stress}(a). Far upstream from the stagnation point, the extensional nature of the flow vanishes, and the polymer stresses relax.

\begin{figure}[t!]
\begin{eqnarray}
\includegraphics[width=0.31\columnwidth,valign=c]{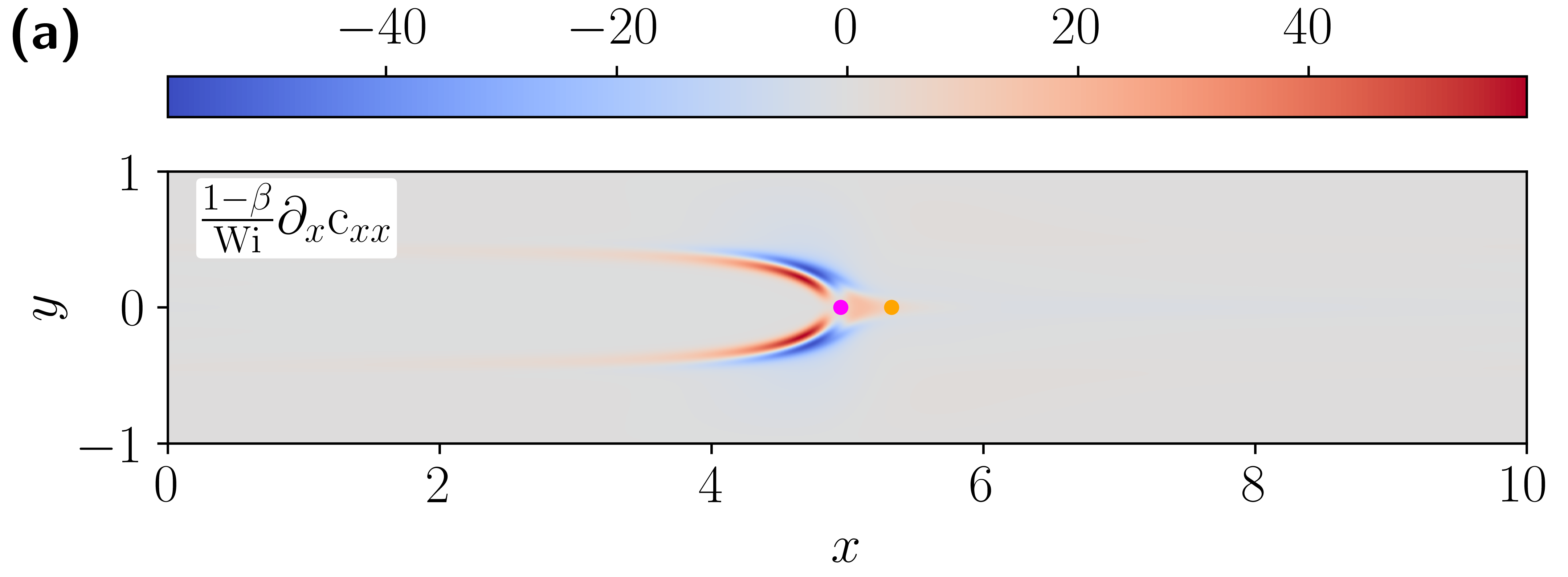} + 
\includegraphics[width=0.31\columnwidth,valign=c]{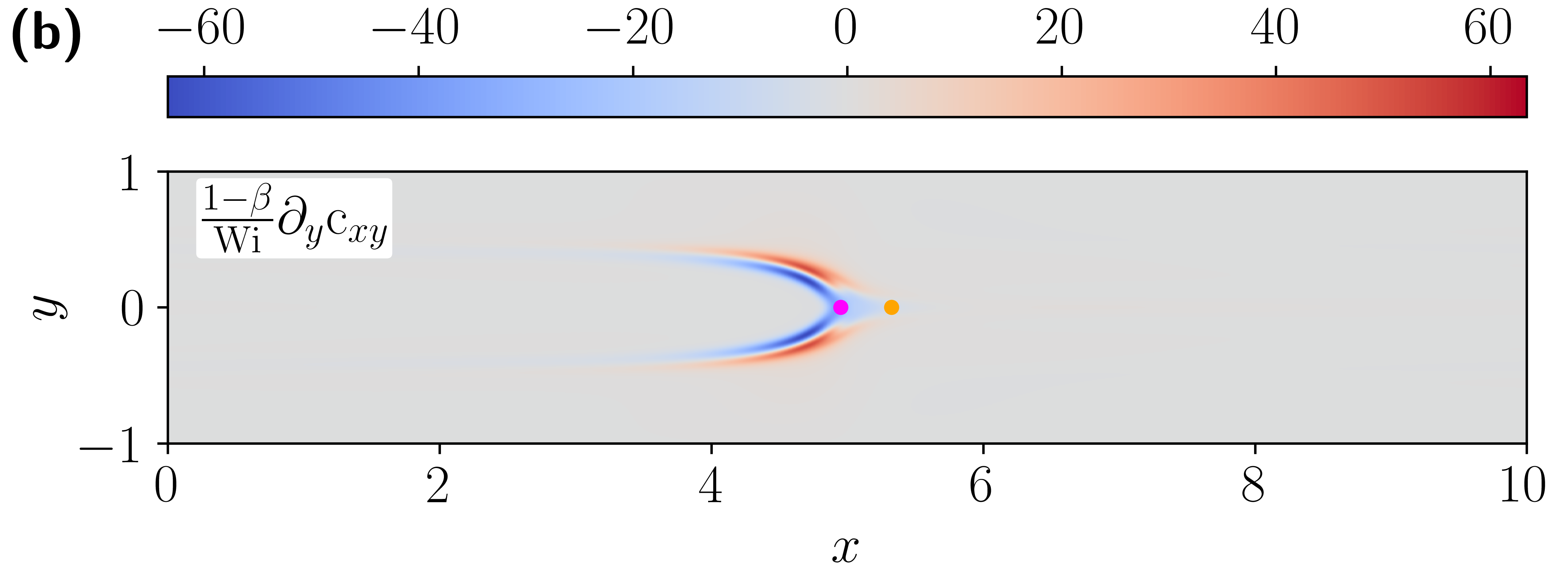} =
\includegraphics[width=0.31\columnwidth,valign=c]{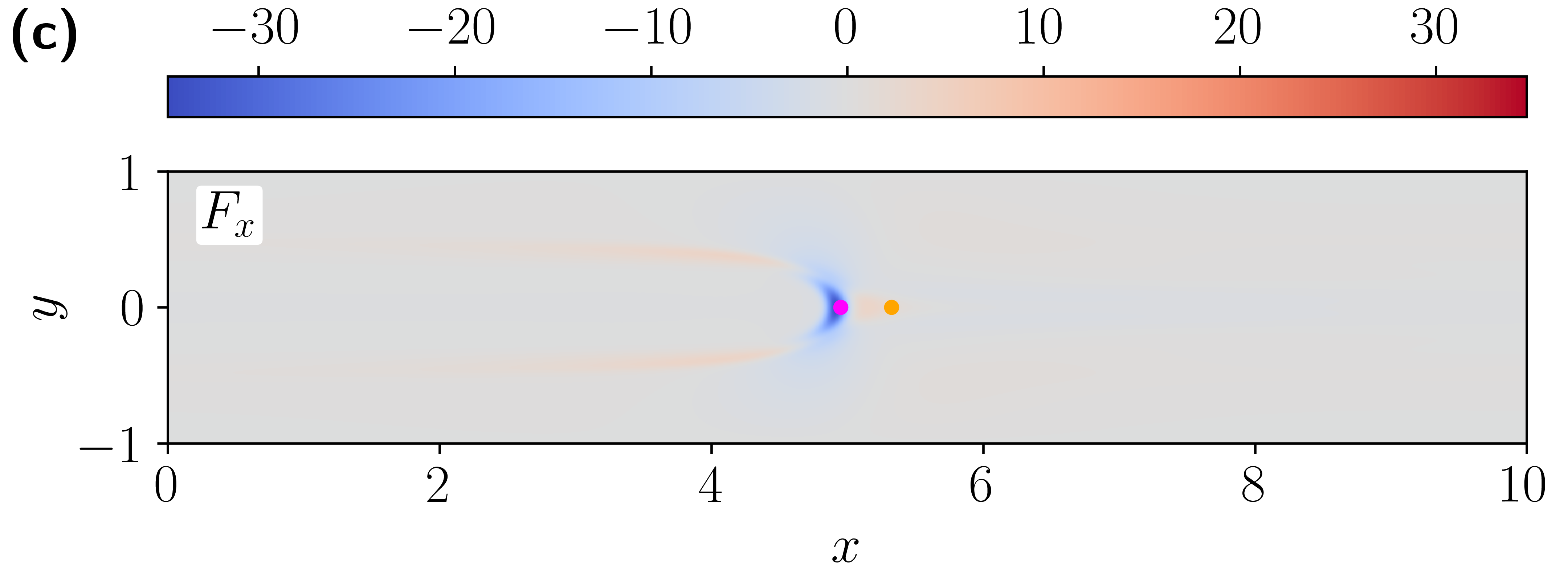} \nonumber \\
\includegraphics[width=0.31\columnwidth,valign=c]{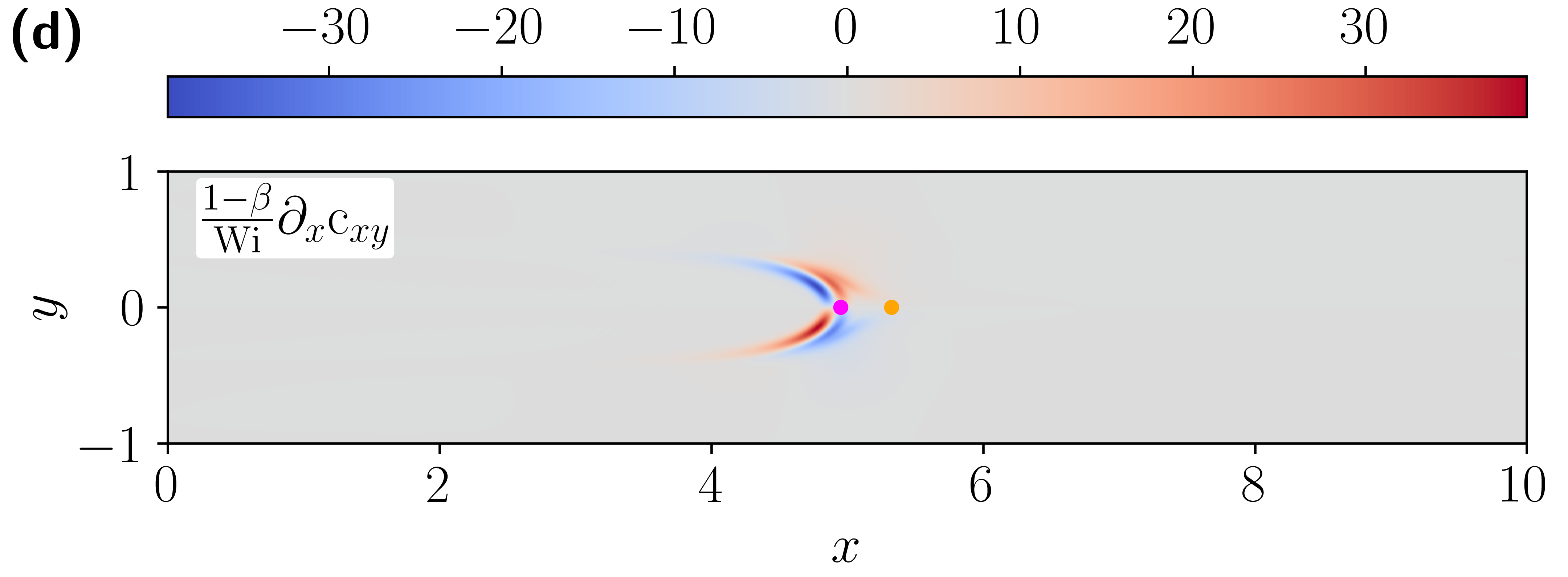} + 
\includegraphics[width=0.31\columnwidth,valign=c]{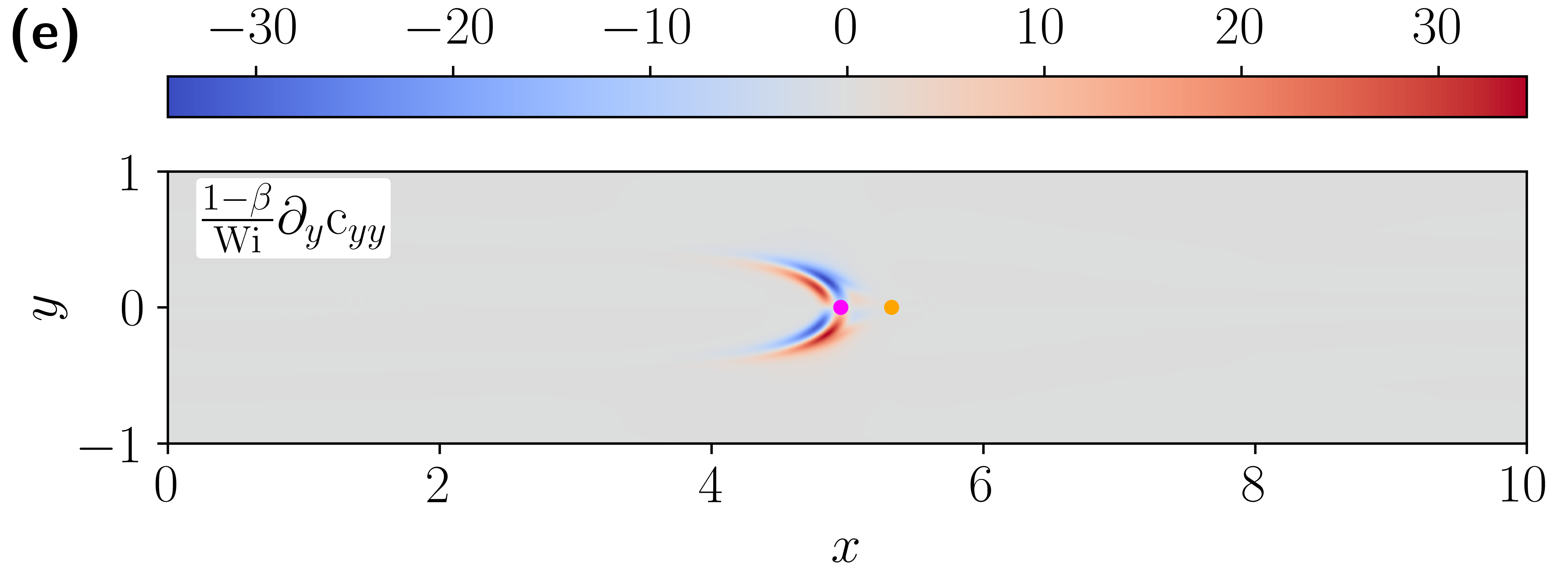} =
\includegraphics[width=0.31\columnwidth,valign=c]{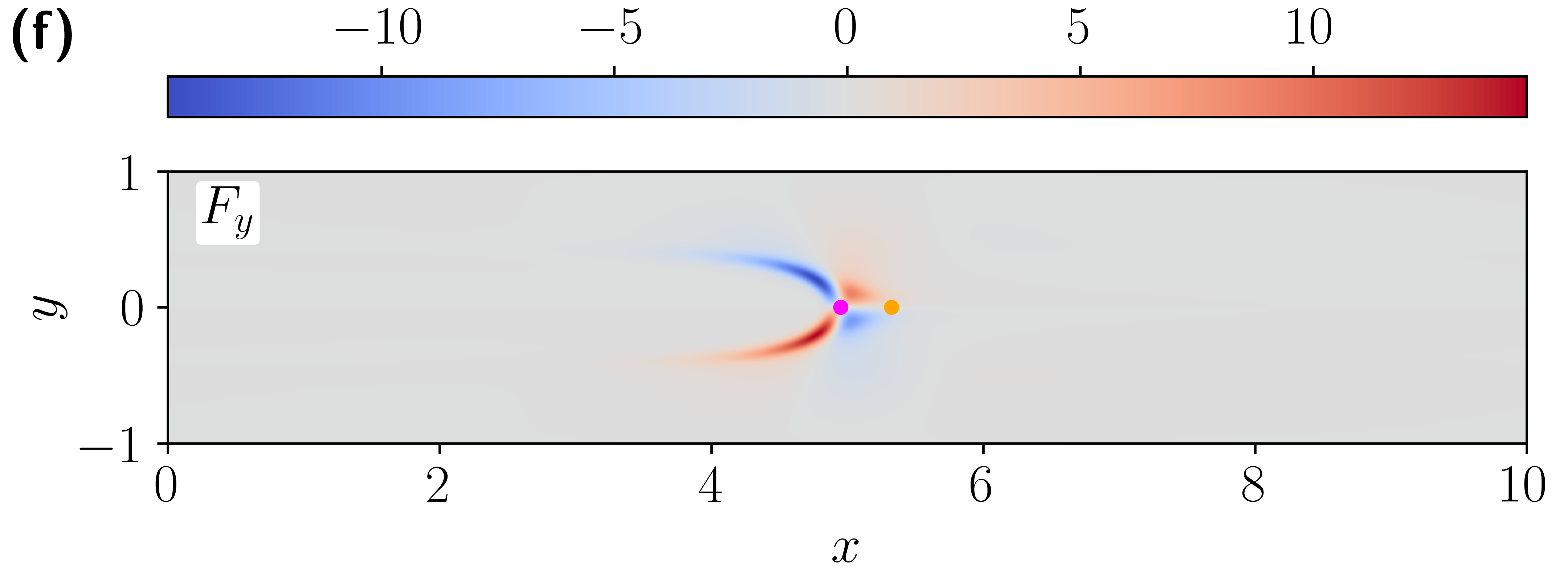} \nonumber
\end{eqnarray}
\caption{Schematic balance of individual terms contributing to the $x$- and $y$-components of the polymer forcing applied to the fluid in the narwhal state presented in Fig.\ref{fig_explanation_velocity}. See text for further explanation.}
\label{fig_explanation_force}
\end{figure}

Until now, the presence of the solitary pair of vortices was taken for granted and we simply worked out the consequences of its existence. To close the self-sustaining cycle, one needs to demonstrate how the spatial stress distribution established above conspires to drive such a secondary flow in the first place. This process is governed by Stokes' equation, 
\begin{align}
\label{stokes_x}
& -\partial_x \delta p + \beta \nabla^2 \delta v_x + \frac{1-\beta}{\Wi} \left[ \partial_x \delta c_{xx} +  \partial_y \delta c_{xy} \right] = 0, \\
\label{stokes_y}
& -\partial_y \delta p + \beta \nabla^2 \delta v_y + \frac{1-\beta}{\Wi} \left[ \partial_x \delta c_{xy} +  \partial_y \delta c_{yy} \right] = 0, \\
\label{incomp}
& \qquad\qquad\qquad \partial_x \delta v_x + \partial_y \delta v_y = 0,
\end{align}
where $\delta {\bm v}$, $\delta {\bm c}$, and $\delta p$ are the velocity, conformation tensor, and the pressure deviations from their laminar profiles, respectively. Based on the conformation tensor profiles associated with the narwhal state, Fig. \ref{fig_explanation_stress}, one might be tempted to argue that since $c_{xx}$ is the largest component of $\bm c$, then $\partial_x c_{xx}$ is the dominant driving term in Eqs.\eqref{stokes_x}-\eqref{incomp}. (Note, that the distinction between the full conformation tensor, $\bm c$, and its deviation from the laminar state, $\delta \bm c$, vanishes around the centreline, since the laminar stresses are only significant in the vicinity of the channel walls.)
Seen as the forcing applied to the fluid, $\partial_x c_{xx}$ points in the downstream direction along the narwhal body, broadly consistent with the velocity field presented in Fig.\ref{fig_explanation_velocity}(a). We now show, however, that the actual force balance is more subtle than that. To this end, we consider the total force density $\left(F_x,F_y\right) =  \frac{1-\beta}{\Wi}\left( \partial_x \delta c_{xx} +  \partial_y \delta c_{xy} , \partial_x \delta c_{xy} +  \partial_y \delta c_{yy} \right)$ applied to the fluid by the polymers. In Fig.\ref{fig_explanation_force} we plot the spatial profile of the force density components $F_x$ and $F_y$ and their individual contributions. As can be seen from there, both force density components comprise two large, and opposite in sign, contributions that nearly cancel each other, resulting in $F_x$ and $F_y$ being significantly smaller in magnitude than their individual terms. As we can also see from Fig.\ref{fig_explanation_force}, $F_x$ is the largest component of the forcing, and we, therefore, conclude that the secondary vortical flow shown in Fig.\ref{fig_explanation_velocity}(a) is driven by the gradients of polymer stresses acting in the negative $x$-direction around the downstream (body) stagnation point. Neither the finely tuned balance of the individual terms in $F_x$ and $F_y$, nor the spatial position where the dominant driving of the fluid is taking place are intuitively obvious, and any quantitative theory of the narwhal state would have to explain both of these observations. 

In January 2025, we have organised a workshop `Chaotic flows in polymer solutions' that was held at the International Centre for Mathematical Sciences in Edinburgh \cite{Chaoflops}. There, Prof. Vincent Terrapon (University of Li\`{e}ge) and their colleagues presented another form of the argument relating the narwhal stress distribution to the solitary vortex pair. Their argument is constructed in a frame oriented locally along the principal components of $\bm c$, and is more concise and elegant than the one presented above \cite{Terrapon2025preprint}.

\subsection{How long is a narwhal?}
\label{subsec:length}

The mechanism presented in Section \ref{subsec:explanation} depicts narwhals as solitary structures: Each narwhal is associated with a pair of vortices and a pair of stagnation points occupying a finite region in the streamwise direction. Moreover, the polymer stresses associated with the narwhal body and its tusk relax along the $x$-direction away from the stagnation points. Since polymer stresses in an unforced fluid decay in time as $\sim\exp{\left(-t/\Wi\right)}$, and since narwhals are travelling-wave solutions and are thus stationary in the co-moving frame, we expect this relaxation to be exponential in $x$, with the typical lengthscale being set by the Weissenberg number.
This argument suggests that narwhals are spatially localised coherent states and should have a typical streamwise size. Here, we assess the validity of this assumption by systematically increasing the streamwise extent of the simulation box $L_x$ and measuring the narwhal length. 

To quantify the latter, we introduce
\begin{align}
\ell = \Big\vert \max(S) - \min(S) \Big\vert, \quad \text{where} \quad S = \Bigg\{x\, \Bigg\vert \, \frac{ \int_{-1}^{+1} dy c_{xx}(x, y)}{\int_{-1}^{+1} dy c_{xx, lam}(y)} - 1 \ge 0.05 \Bigg\}.
\label{eq:2D_embeddings.outer_length_scale}
\end{align}
This definition identifies a narwhal as a streamwise interval $S$ where $c_{xx}$ is at least $5\%$ larger than its laminar value (averaged across the gap); the normalisation with the laminar profile ensures that we can compare structures across different values of $\Wi$. We note that the threshold value of $5\%$ is rather arbitrary, and we have confirmed that the same qualitative conclusion holds for other, sufficiently small, threshold values.

\begin{figure}[t!]
\centering
\includegraphics[width=1\columnwidth]{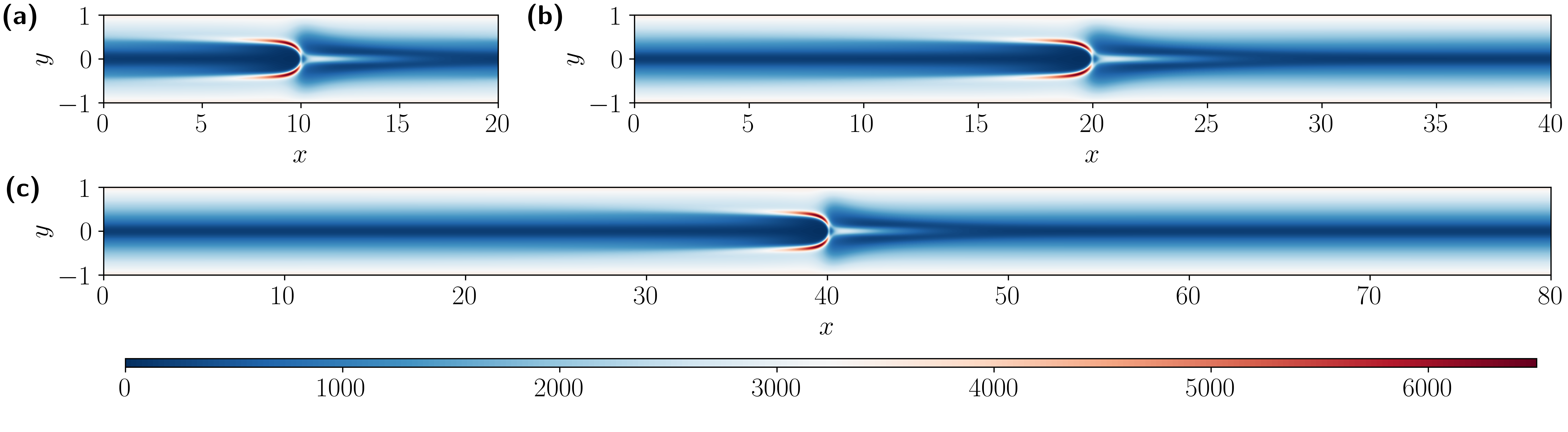}\\ \quad \\
\includegraphics[width=0.35\columnwidth]{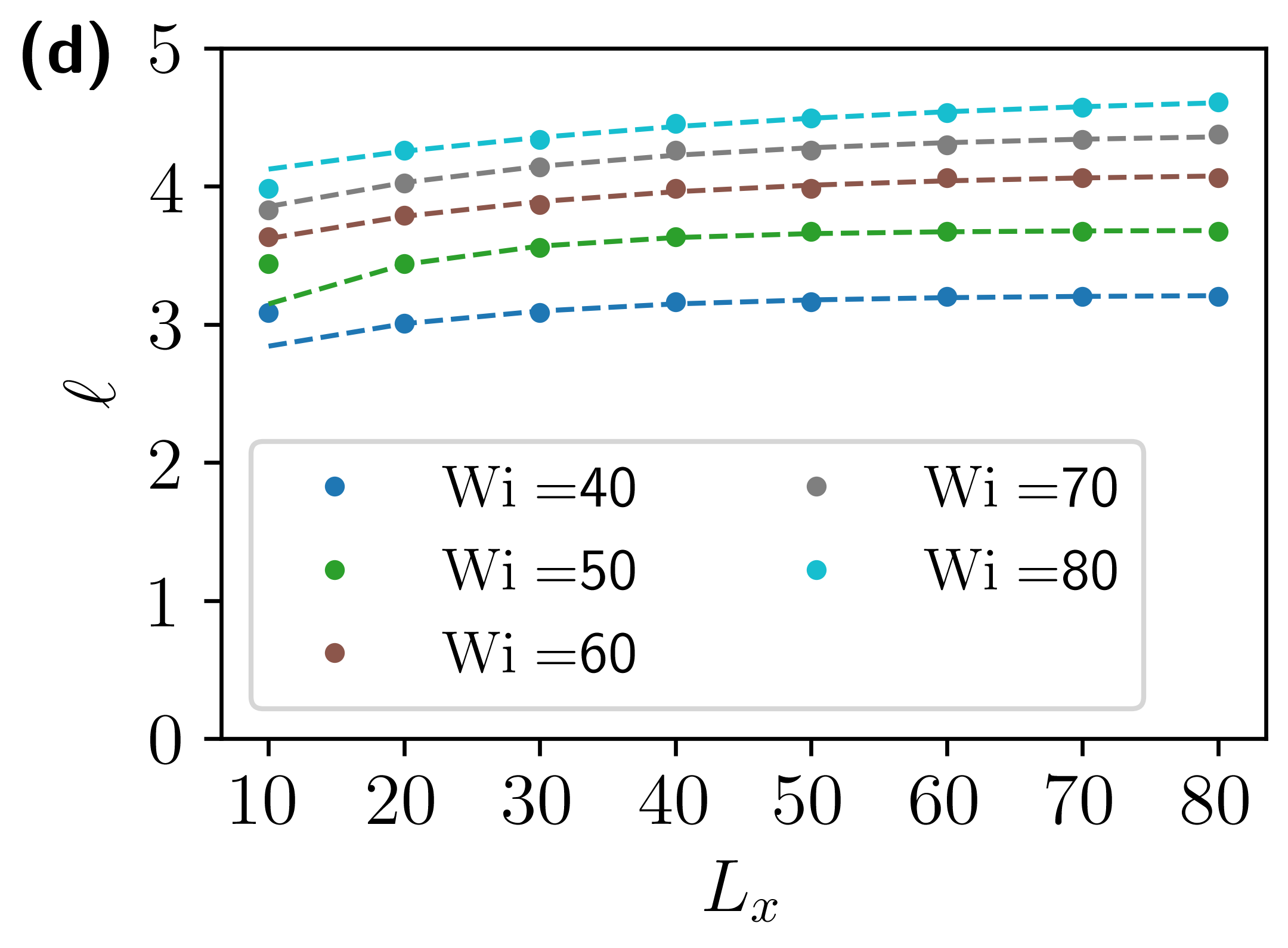}
\includegraphics[width=0.35\columnwidth]{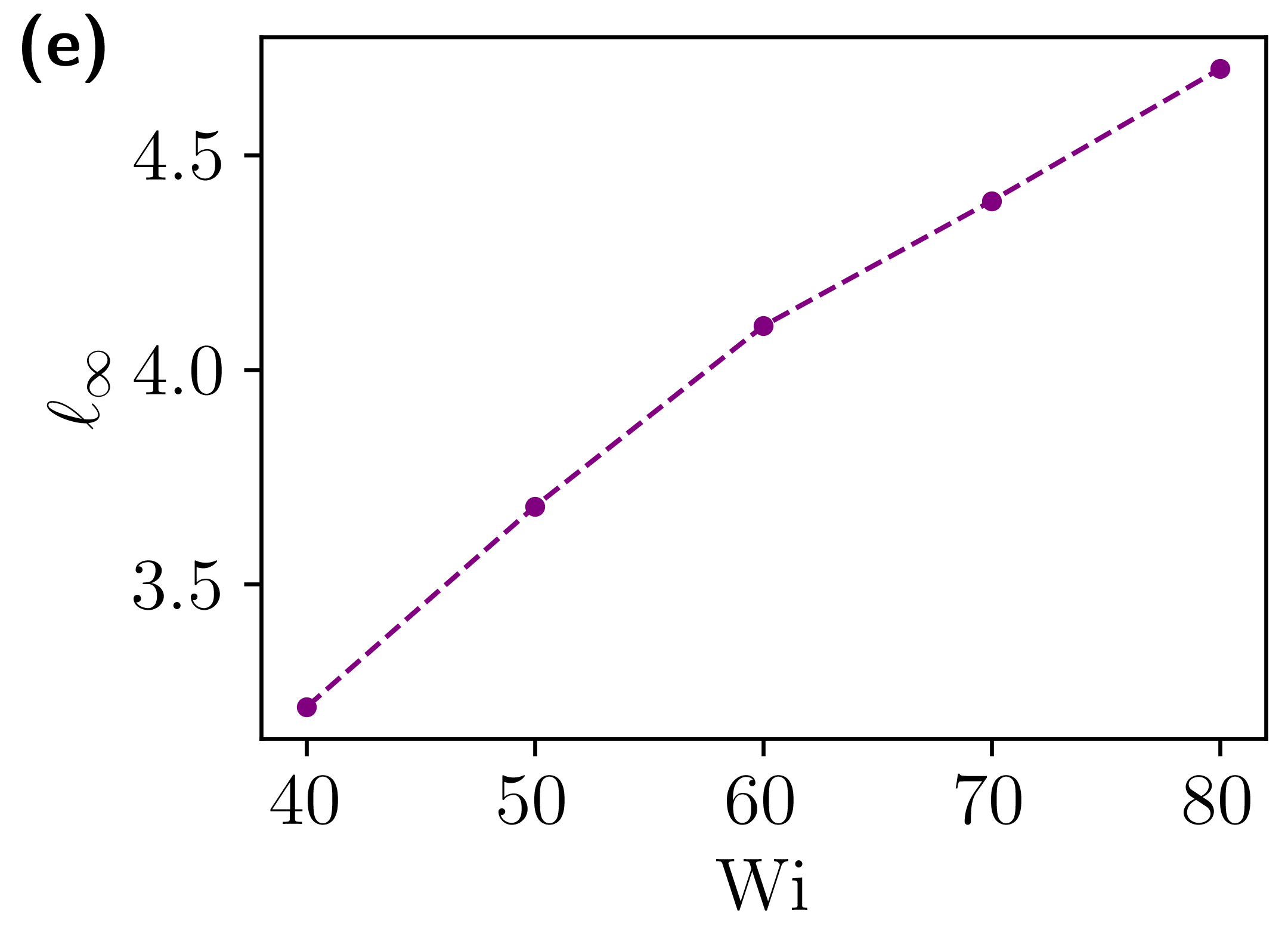}
\caption{Narwhals in long boxes for $\beta=0.8$. (a)-(c) Sample configurations of polymer stretch for $\Wi=80$ with (a) $L_x=20$, (b) $L_x=40$, and (c) $L_x=80$. (d) The spatial lengthscale $\ell$ measured as a function of $L_x$ for various values of $\Wi$. (e) Fitted values of the asymptotic lengthscale $\ell_\infty$ as a function of $\Wi$.}
\label{Lx}
\end{figure}

Next, we simulate single narwhals states at various values of $\Wi$ while varying the domain length $L_x$, see Fig.\ref{Lx}(a)-(c), for example. In Fig.\ref{Lx}(d), we plot the spatial lengthscale $\ell$ associated with these states and observe that for all values of $\Wi$ the narwhal length $\ell$ appears to increase with the streamwise domain size and then saturate at a value independent of $L_x$. To extract the asymptotic value of the length, $\ell_\infty$, we fit the data in Fig.\ref{Lx}(d) to a phenomenological expression $\ell(L_x) = \ell_\infty - A \exp\left(-L_x/B\right)$, where $A$ and $B$ are constants. In Fig.\ref{Lx}(e), we plot $\ell_\infty$ thus extracted as a function of $\Wi$ and
observe that it is increasing with $\Wi$. While this behaviour is broadly in line with the argument presented above that relates the spatial extent of a narwhal with stress relaxation in the absence of any forcing, neither the values of $\ell_\infty$, nor their dependence on $\Wi$ are currently understood.

\begin{figure}[t!]
\centering
\includegraphics[width=0.7\columnwidth]{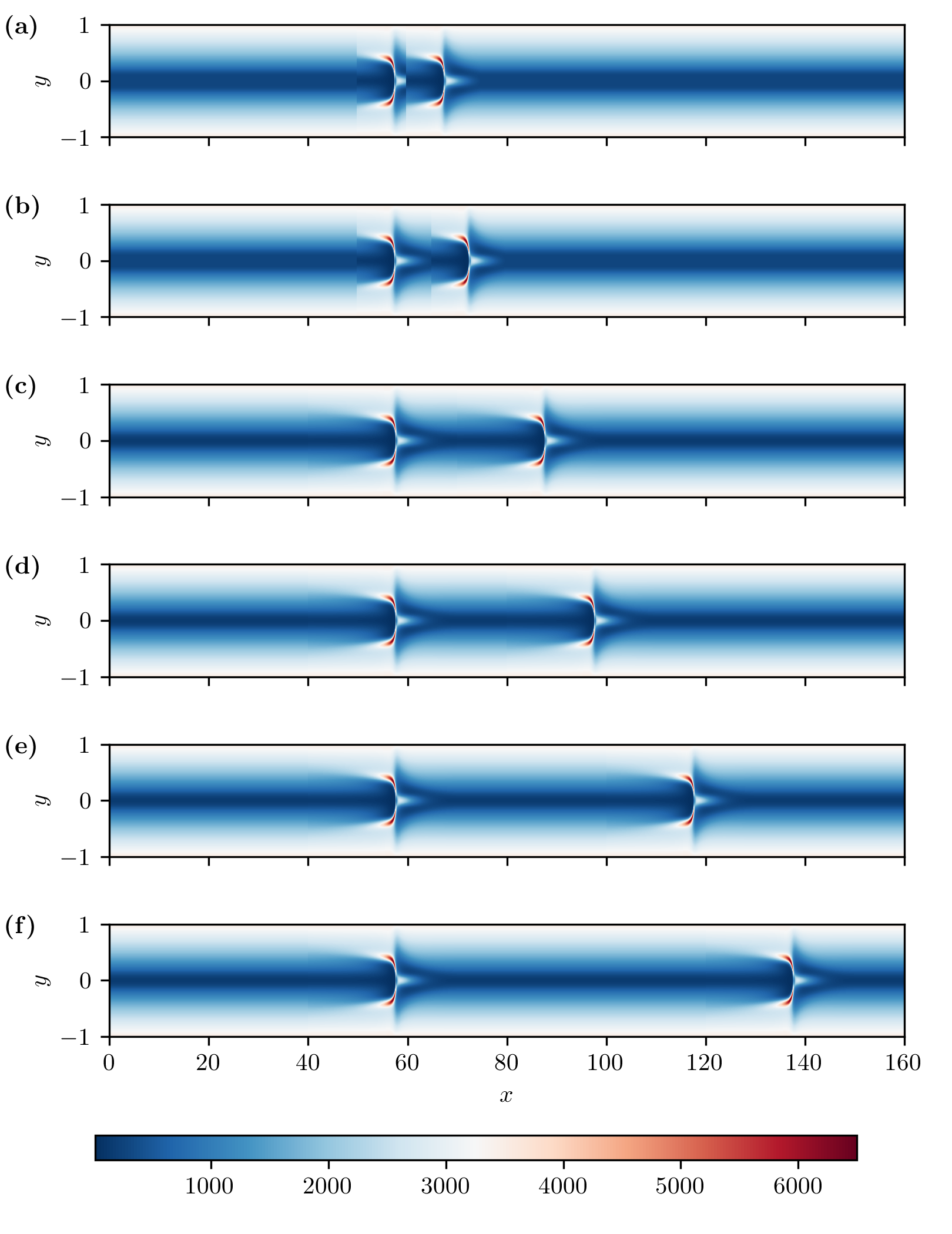}
\caption{The initial conditions at $\beta=0.8$ and $\Wi=80$ constructed by placing two narwhal states at an initial distance $s(0)$ from each other. Panels (a)-(f) correspond to $s(0)=\left\{10,15,30,40,60,80\right\}$, respectively. Visible discontinuities in the polymer stretch profiles are the artefacts of the construction procedure; they relax on a very short timescale.}
\label{two_narwhals_init}
\end{figure}

\subsection{Interactions between narwhals}
\label{subsec:interactions}

Another way to assess the typical streamwise extent of a narwhal is to study the interaction between two of such states. Within this approach, the minimum separation at which narwhals are not affected by each other's presence defines the lengthscale of a solitary structure. It also allows us to assess the type of interactions between narwhals within a blessing. Here, we present some initial results of such a study.

The initial conditions we used are presented in Fig.\ref{two_narwhals_init}. To construct them, we took a converged narwhal state in a short box and embedded it at two positions in a large domain, separated by an initial distance $s(0)$. We then monitored how the narwhal separation $s(t)$ changed in time. The details of the embedding procedure and the definition of $s$ can be found elsewhere \cite{LellepThesis}. 

\begin{figure}[t!]
\centering
\includegraphics[width=0.45\columnwidth]{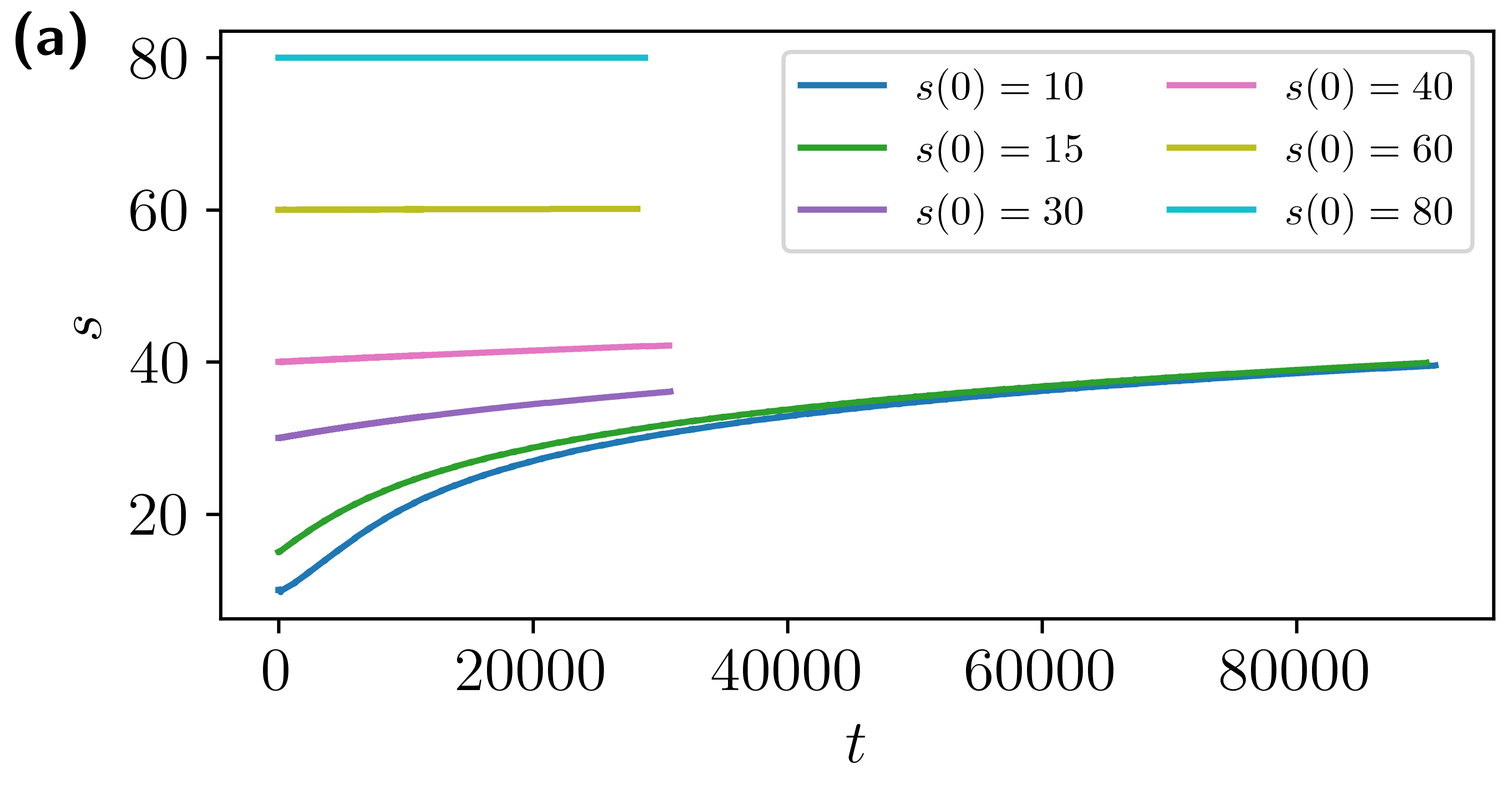}
\includegraphics[width=0.45\columnwidth]{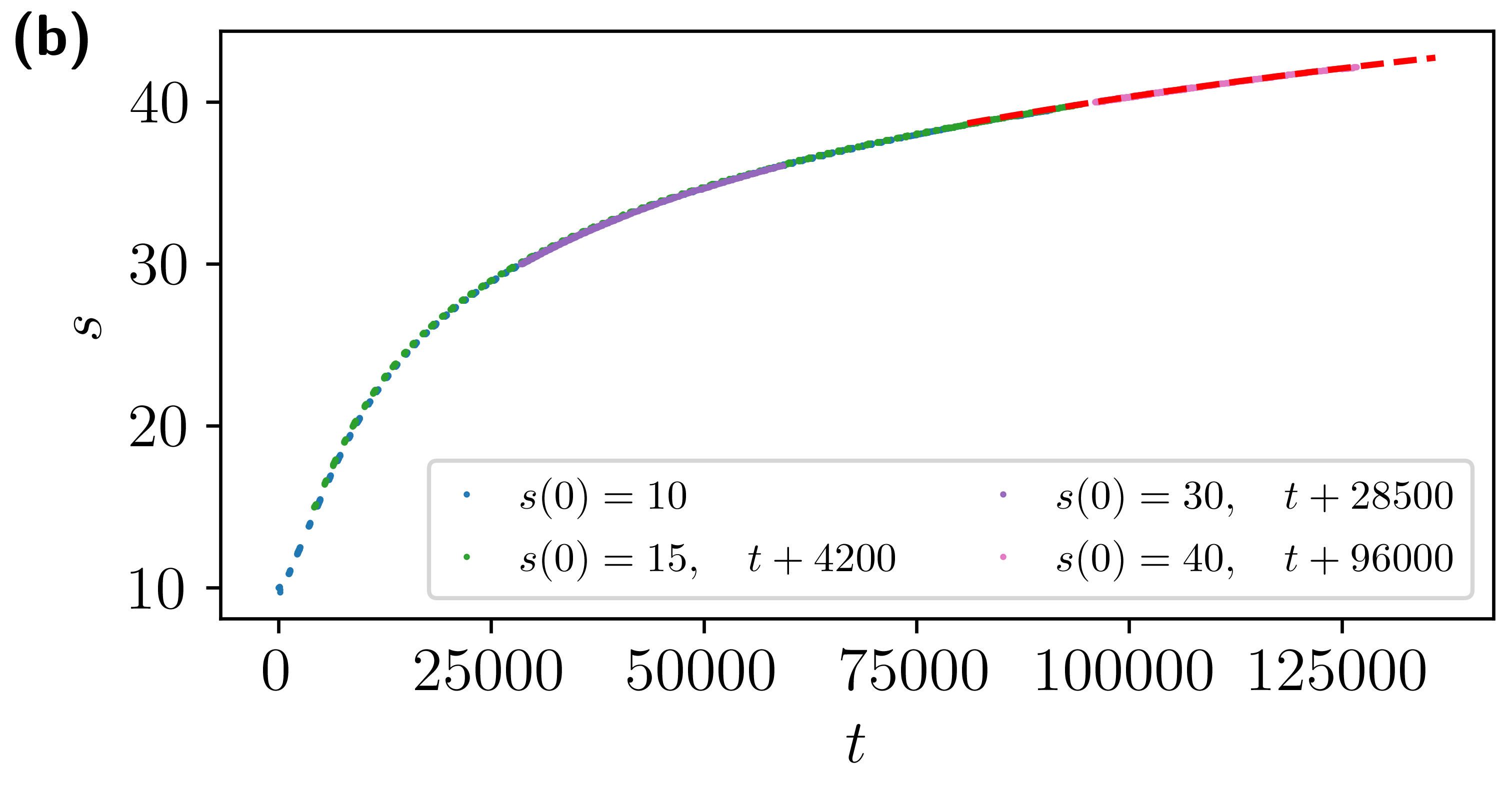}
\caption{The narwhal separation as a function of time for various values of the initial separation $s(0)$. (a) The original data and (b) the data from panel (a) shifted in time as indicated.}
\label{two_narwhals_s_of_t}
\end{figure}

For sufficiently large initial separations, $s(0)=60$ and $s(0)=80$, we observe no measurable change in the narwhal separation, see Fig.\ref{two_narwhals_s_of_t}(a), indicating that these initial distances are larger than a typical interaction range. For smaller initial separations, $s(t)$ is increasing in time, indicating that two narwhal states repel each other. This behaviour is similar to the interaction between turbulent puffs in Newtonian pipe flow that were shown to increase their mutual separation after splitting events \cite{Shimizu2014,Barkley2016,Avila2023}. We now show that this interaction does not depend on the history of the narwhal co-evolution, and is simply a function of their instantaneous separation. To this end, we add time shifts to individual $s(t)$ curves to obtain a single master curve, common for all initial separations $s(0)<60$, see Fig.\ref{two_narwhals_s_of_t}(b). The existence of such a master curve implies that two narwhals separated by a distance $s_*$ proceed to interact in the same way regardless whether we started a new simulation with $s(0)=s_*$ or whether a simulation with a different $s(0)$ evolved to have $s_*$ as its current separation. 

As can be seen from Fig.\ref{two_narwhals_s_of_t}, the minimum separation at which narwhals are not affected by each other's presence is larger than $s=40$ but smaller than $s=60$. We note that the longest time series presented in Fig.\ref{two_narwhals_s_of_t} were obtained in a simulation performed on $1024$ cores for $1080$ hours. We found it to be impractical to obtain the plateau value of the master curve in Fig.\ref{two_narwhals_s_of_t}(b) by extending these runs even further. Instead, we fit the last portion of the master curve to a phenomenological expression $s(t) = s_\infty - A \exp\left(-B\,t\right)$ and obtain $s_\infty \approx 49$, with the fit shown in Fig.\ref{two_narwhals_s_of_t}(b) by the red dashed line. 

The estimated plateau value is an order of magnitude larger than the narwhal length $\ell_\infty$ measured in Sec.\ref{subsec:length}. This discrepancy might originate in the different nature of the measurement performed there: while $\ell_\infty$ estimates the typical decay length of the narwhal stresses, $s_\infty$ describes the distance at which the velocity fields created by the narwhals are too small to advect each other significantly. In the absence of inertia, the far-field behaviour of such velocity fields is algebraic \cite{kim2013microhydrodynamics} and it is conceivable that their mutual presence is felt at any spatial separation, however large. In such scenario, our observations of narwhals not changing their separation for $s(0)=60$ and $s(0)=80$ might be an artefact of the finite box size and simulation time, with the true value of $s_\infty$ diverging in the thermodynamic limit. Currently, the correct value of $s_\infty$ remains an open question. 

\begin{figure}[t!]
\centering
\includegraphics[width=1\columnwidth]{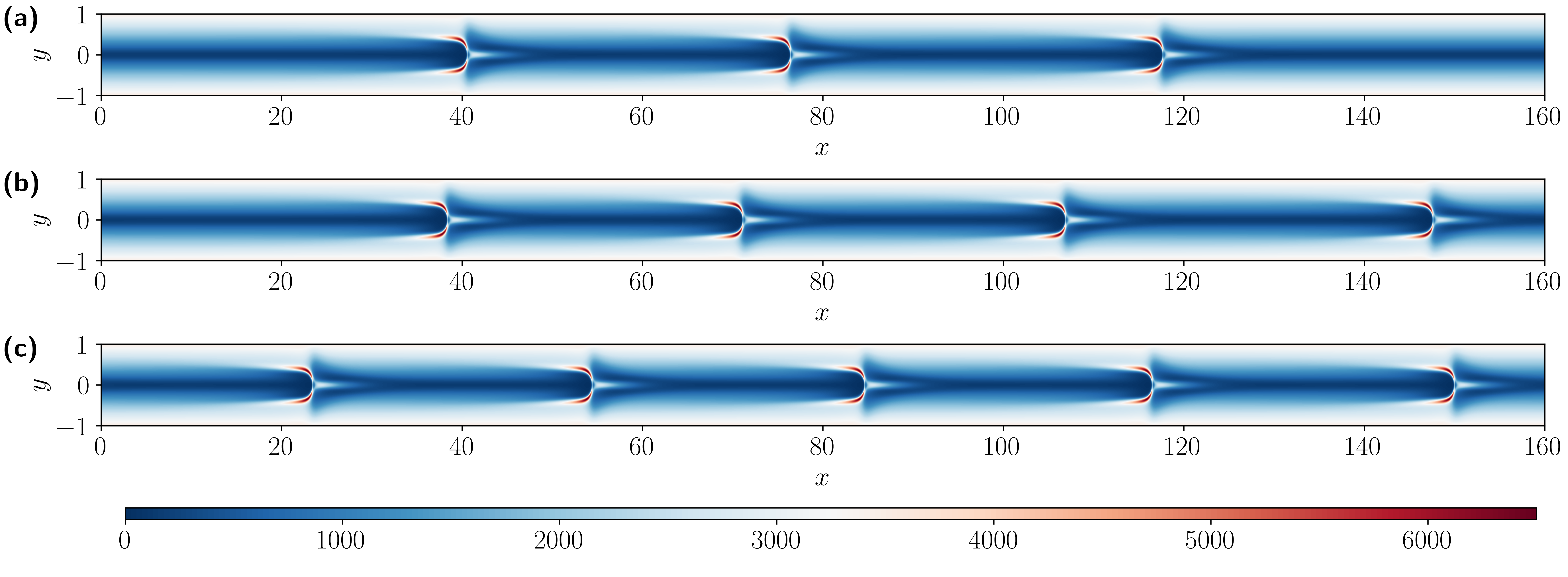}
\caption{Examples of blessings with (a) three, (b) four, and (c) five narwhals in a box with $L_x=160$, $\beta=0.8$, and $\Wi=80$.}
\label{N345}
\end{figure}

We conclude by mentioning here that the behaviour of a narwhal blessing becomes significantly more complicated when more than two narwhals are present in the domain. In Fig.\ref{N345}, we present examples of blessings comprising three, four, and five narwhals, respectively, and observe that their inter-narwhal separations depend on the exact number of narwhals in the blessing; they also decrease towards the trailing end of the latter. Also, when a blessing is confined to a simulation box with a sufficiently small $L_x$, the ensuing dynamics can artificially become time-dependent and, perhaps, even chaotic \cite{LellepThesis}, thus stressing the need of carefully accessing the lengths of simulation boxes used to study elastic turbulence.

\subsection{Are two-dimensional narwhals relevant for elastic turbulence?}
\label{subsec:lininstab}

Until now, we have carefully avoided perhaps the most pertinent question: Are narwhals experimentally relevant? Can they be observed in actual realisations of pressure-driven flows of dilute polymer solutions? At face value, the results presented above suggest that this is unlikely. There are two main reasons for this conclusion.

Firstly, narwhal states are travelling-wave solutions that move with constant speed down the channel. As such, they are steady in a co-moving frame, and all associated global observables remain constant in time. In contrast, experiments consistently indicate that purely elastic parallel shear flows undergo a direct transition to elastic turbulence, with all observables fluctuating in time~\cite{Bonn2011,Pan2013,Qin2017,Qin2019,Steinberg2022,Shnapp2022,Li2023,Li2024,Li2024a,Li2025}. Secondly, narwhal states are intrinsically two-dimensional.

It is therefore plausible that their status is similar to that of the Newtonian Tollmien-Schlichting waves \cite{SchmidHenningson2000}. These are exact two-dimensional travelling-wave solutions of the Navier-Stokes equations, discovered in pressure-driven channel flows \cite{Herbert1976,Jimenez1990}. They arise from a linear instability at $\Re=5772$, but exist subcritically above $\Re \gtrapprox 2700$. At sufficiently high values of $\Re$, Tollmien-Schlichting waves themselves become unstable and transition to chaos in two-dimensional channel-flow simulations \cite{Jimenez1990}. Their behaviour in three dimensions, however, is very different: when embedded in 3D domains, two-dimensional Tollmien-Schlichting waves lose their stability and evolve into very different, fully three-dimensional states \cite{Orszag1983}.

To examine whether a similar scenario applies to narwhal states, in Lellep \emph{et al.}\cite{Lellep2023} we studied their linear stability in three dimensions. To this end, a two-dimensional narwhal state, translationally invariant along the spanwise direction, was embedded in a three-dimensional simulation box. The resulting state was perturbed by adding a minute amount of Gaussian noise with amplitude $O(10^{-7})$ relative to the largest stress of the narwhal solution. To assess stability, the equations of motion, Eqs.\eqref{eq:ptt}-\eqref{eq:incomp}, were linearised, and their time evolution was followed.

In all cases considered, two-dimensional narwhal states were found to be linearly unstable when embedded in three-dimensional domains, with the typical lengthscale of the instability being comparable to the channel width. This conclusion was verified across a wide range of $\beta$, $\Wi$, and $\Re$, including the cases studied in Morozov \cite{Morozov2022} and Page \emph{et al.}~\cite{Page2020}.

One might therefore conclude that two-dimensional narwhals are an academic curiosity, dynamically irrelevant in three spatial dimensions. In the next Section, we continue this line of investigation by following the linear instability discussed here until it saturates into a fully three-dimensional nonlinear state. As we show there, the resulting state exhibits strong spatio-temporal intermittency, sharing many features with experimental observations of elastic turbulence. Interestingly, instantaneous stress profiles in the streamwise -- wall-normal cross-section often resemble those of two-dimensional narwhals, suggesting that although linearly unstable, narwhal solutions remain dynamically relevant by organising the chaotic dynamics of three-dimensional flows.

\newpage
\section{Elastic turbulence in three dimensions}
\label{pnassection}

Motivated by the observation that the two-dimensional narwhals become linearly unstable when embedded in three-dimensional domains (see \cite{Lellep2023} and Section \ref{subsec:lininstab} above), here we explore the non-linear states that emerge after the instability. Unless explicitly stated, all runs in this Section employed $L_x = L_z=10$, $\Re=0.01$, $\beta=0.8$, $\epsilon = 10^{-3}$ and $\kappa = 5\cdot 10^{-5}$. We have verified that the base flow is linearly stable for all parameters considered here. The results presented below were originally reported in \cite{Lellep2024}, with the exception of the run in Section \ref{sect:longbox}. 

To study the non-linear stability of the three-dimensional polymeric channel flow, we performed a series of direct numerical simulations varying $\Wi$ from $0$ to $150$. We employed three types of initial conditions. Firstly, we have used small-amplitude Gaussian noise in $c_{xx}$ added to the two-dimensional narwhals copied along the spanwise direction, as discussed in Section \ref{subsec:lininstab}. Secondly, after we have successfully simulated elastic turbulence at some $\Wi$, we used random snapshots sampled in the statistically steady-state to initialise other simulations at a different value of $\Wi$. Finally, we have perturbed the laminar state by adding a finite-amplitude Gaussian noise to the $c_{xx}$ component of the conformation tensor. As we discuss below, the latter protocol resulted in runs that took a long time to reach elastic turbulence and we have performed a relatively small number of such simulations.

\begin{figure}[t!]
\centering
\includegraphics[width=0.8\columnwidth]{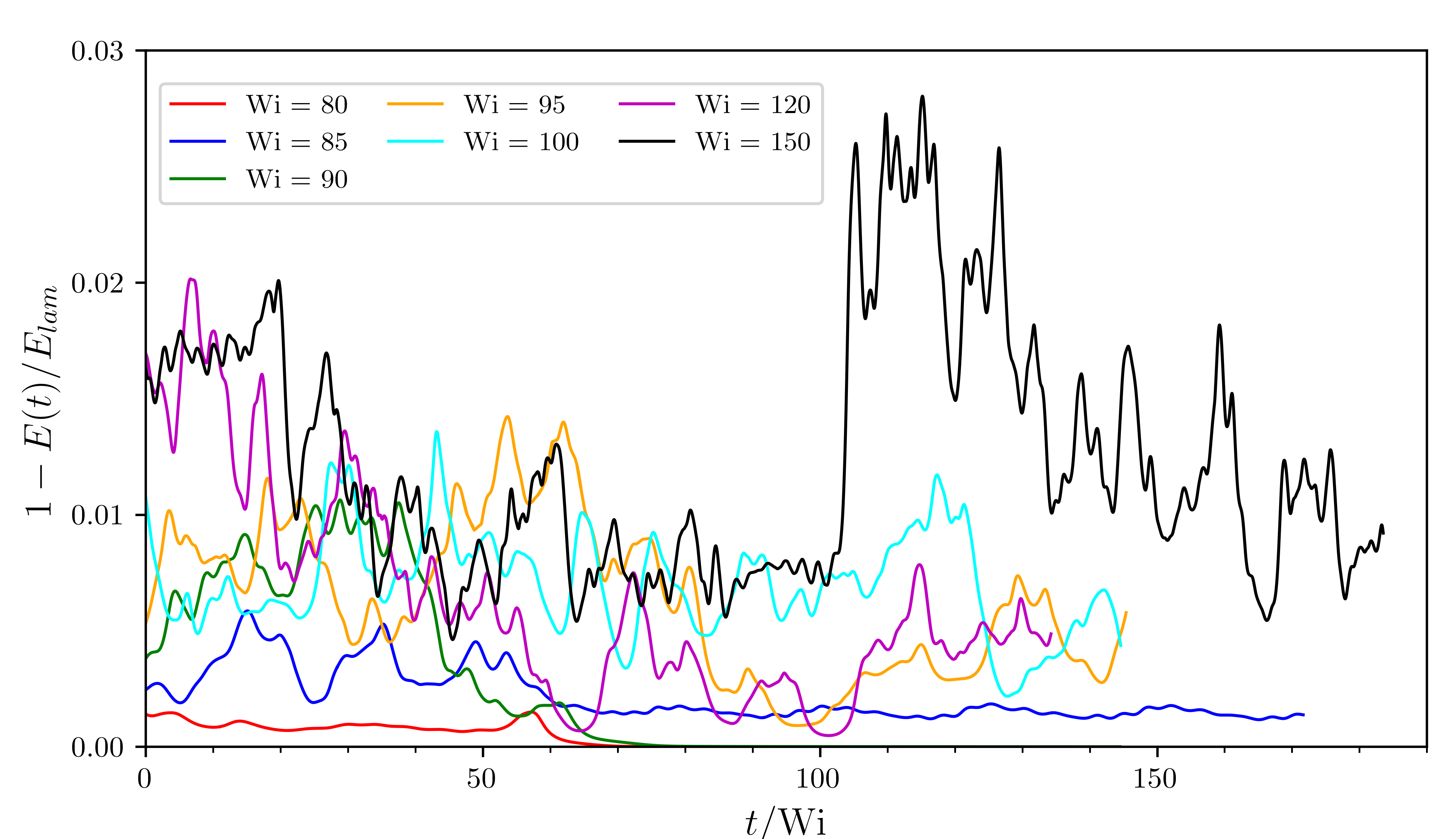}
\caption{Temporal evolution of the reduced kinetic energy for various values of $\Wi$. Replotted from the data originally presented in Fig.1C of Lellep \emph{et al.} \cite{Lellep2024}.}
\label{KEvsT}
\end{figure}

\subsection{Localised coherent structures of elastic turbulence}

For all values of $\Wi<80$, the initial conditions always return to the laminar state independent of the strength and type of the original perturbation. Note, that we have never observed a three-dimensional simulation returning to a two-dimensional narwhal state, translationally invariant along the spanwise direction. Above $\Wi\sim 80$, small perturbations decay in time, while sufficiently strong perturbations grow, in line with the bifurcation-from-infinity scenario discussed in Section \ref{sec:intro}. The precise value of the threshold amplitude, separating the growing from the decaying disturbances, depends on the type of perturbation used, similar to the Newtonian case \cite{Peixinho2007}. After the instability sets in, we observe chaotic temporal oscillations of the total kinetic energy, $E(t)$, see Fig.\ref{KEvsT}, concurrent with oscillations of the total polymer stretch (see below). For the values of $\Wi$ close to the onset value $\Wi=80$, we observe long-time chaotic dynamics followed by sudden flow relaminarisation. This process is stochastic in nature, with the relaminarisation time changing unpredictably when the same run is repeated with another random initial condition. Such events are similar to the behaviour observed in transitional parallel shear flows of Newtonian fluids \cite{Avila2011,Barkley2016,Avila2023}. For larger values of $\Wi$, on the other hand, chaotic fluctuations persist for the whole duration of the simulation, spanning many polymer relaxation times (i.e. large values of $t/\Wi$ in our dimensionless units). 

\begin{figure}[t!]
\centering
\includegraphics[width=1\columnwidth]{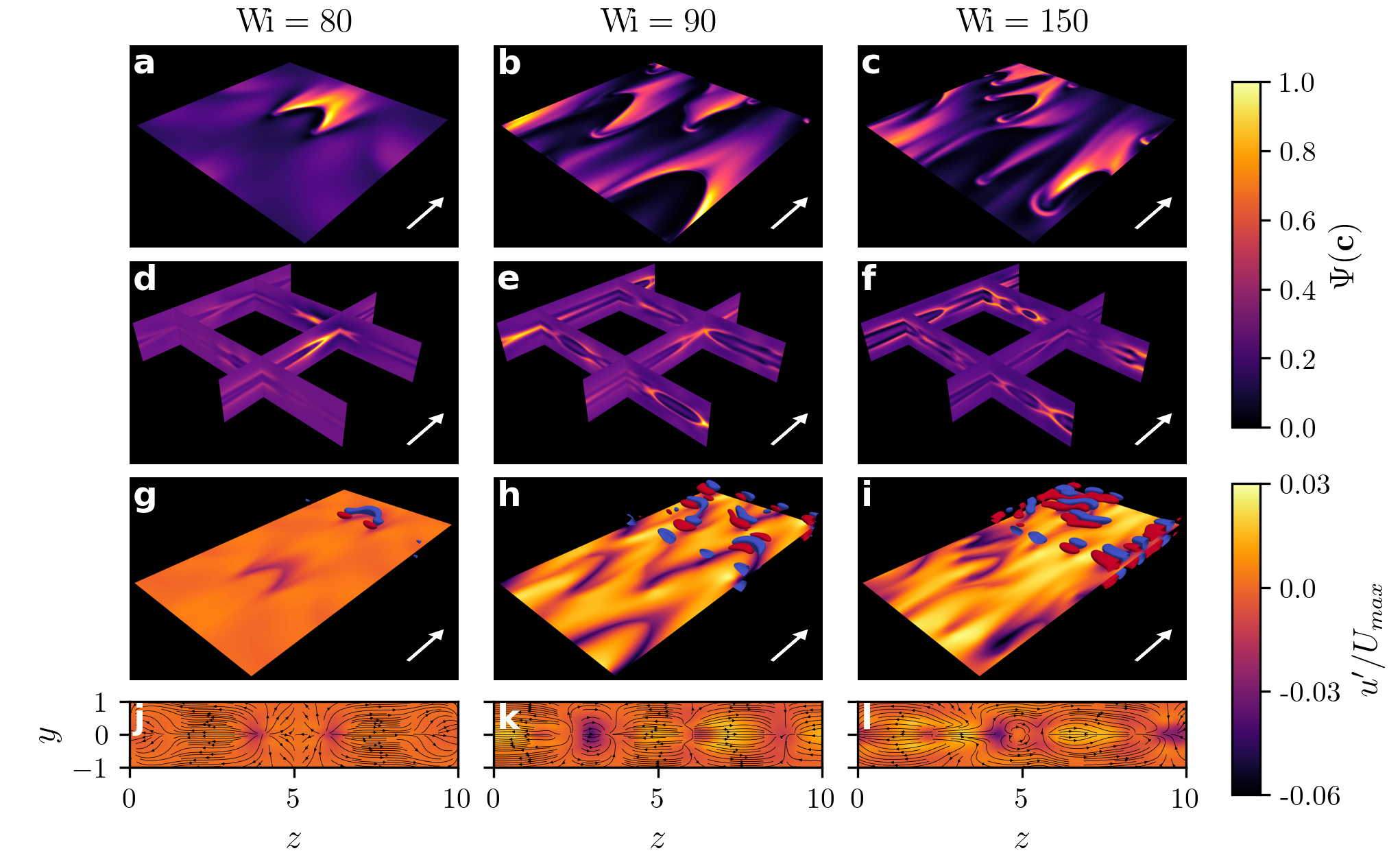}
\caption{Instantaneous spatial structure of elastic turbulence at various $\Wi$. The mean flow direction is indicated by the white arrows. (a)-(c) Midplane profile of the polymer stretch (proportional to the quantity $\Psi({\bm c})$). (d)-(f) Vertical slices at several positions across the channel demonstrating the presence on instantaneous two-dimensional narwhal states. (g)-(i) The midplane profile of the streamwise velocity and the isosurfaces of the Q-criterion identifying the presence of spanwise-oriented vortices. In these panels, the simulation domain is repeated twice in the streamwise direction to improve readability. (j)-(l) Velocity field in the spanwise -- wall-normal plane. The deviation of the streamwise velocity from its mean is given by the colour, while the in-plane velocity components are given by the streamlines. The figure is replotted from the data originally presented in Fig.2 of Lellep \emph{et al.} \cite{Lellep2024}. We refer the interested reader to Ref.\cite{Lellep2024} for the precise definitions of the quantities plotted here.}
\label{pnas_main}
\end{figure}

The main result of our work is shown in Fig.\ref{pnas_main}, where we present the spatial velocity profiles of the total polymer stretch and velocity associated with elastic turbulence in pressure-driven flows. For all $\Wi$ studied here, we observe that the largest deviation of the polymer stress from its laminar proﬁle is localised
in a thin sheet around the channel centreline, see Figs.\ref{pnas_main}(a)-(c), while the flow is essentially laminar close to the walls. This is in a stark contrast with Newtonian turbulence in parallel shear flows, where the strongest ﬂuctuations are found close to the walls \cite{Smits2011}.

\begin{figure}[t!]
\centering
\includegraphics[width=0.35\columnwidth]{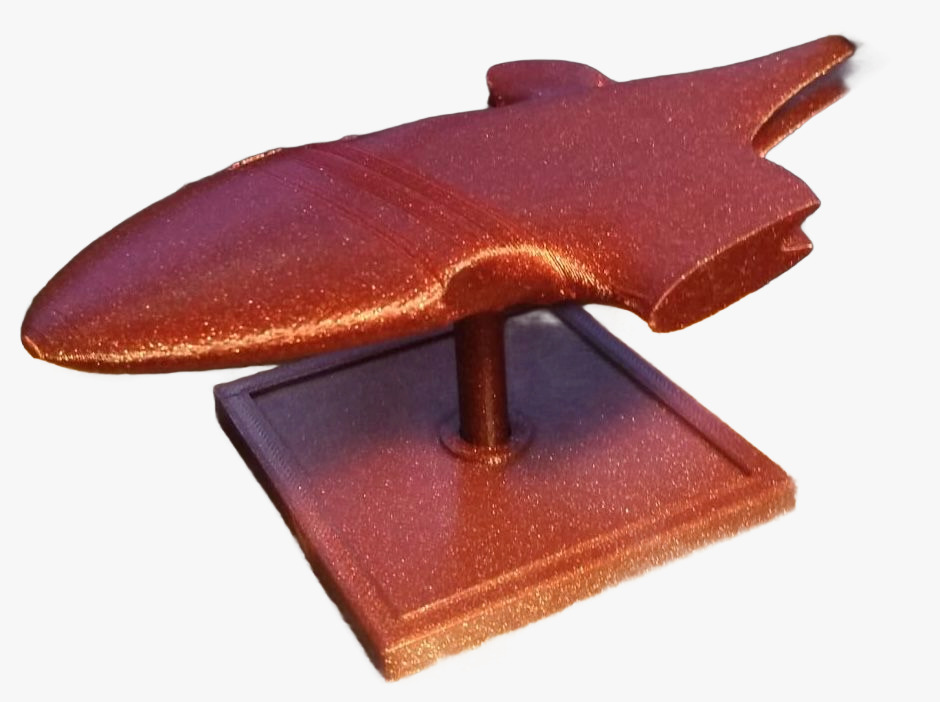}
\caption{An illustration of the spatially localised coherent states of elastic turbulence. As discussed in the text, the isosurface of the polymer stretch was extracted from our simulations and 3D printed to manufacture the object presented here. The supporting stand is not a part of the coherent state. The continuum of tusks in the midplane is too thin to manufacture.}
\label{3dprinted}
\end{figure}

For low values of the Weissenberg number, $\Wi \approx 80$, the polymer extension exhibits a spatially localised profile, reminiscent of turbulent puffs and spots in Newtonian pipe and channel flows \cite{Tuckerman2020,Graham2021}, respectively. At larger $\Wi$, these localised structures proliferate across the domain, undergoing chaotic splitting and merging (see supplementary movies in Lellep \emph{et al.} \cite{Lellep2024}). Throughout these complex dynamics, the stress field in the streamwise -- wall-normal plane, Figs.\ref{pnas_main}(d)-(f), retains the characteristic features of the two-dimensional narwhal 
states discussed in Section \ref{sec:2D}. This suggests that linearly unstable two-dimensional narwhals underpin the chaotic dynamics near the onset of elastic turbulence in three dimensions.

In Fig.\ref{pnas_main}, and in what follows, we focus primarily on the stress distribution in the channel midplane, as this region exhibits the strongest deviation of the total polymer stretch, $\mathrm{Tr}\,\bm c$, from its laminar profile. Nevertheless, Figs.\ref{pnas_main}(d)-(f) show that the three-dimensional localised structures also extend significantly away from the midplane. To better illustrate their spatial form, we extracted an isosurface of $\mathrm{Tr}\,\bm c$ from the data in Fig.\ref{pnas_main}, and produced a 3D-printed representation of the structure, shown in Fig.\ref{3dprinted}. Note that the supporting stand is not part of the structure, and the midplane stresses, corresponding to the continuum formed by the tusks of the two-dimensional narwhals in Figs.\ref{pnas_main}(d)-(f), were too thin to fabricate. The surface shown in Fig.\ref{3dprinted} can be thought of as the locus of points formed by the arrangement of the two-dimensional narwhal bodies visible in Figs.\ref{pnas_main}(d)-(f). The resulting flat, pancake-like structure, localised in all directions, appears to represent a fundamental building block of elastic turbulence in the transition regime.


We now turn to the velocity field associated with the states presented in Fig.\ref{pnas_main}. The most prominent features are found in the streamwise velocity component: similar to the polymer stretch, significant deviations of $v_x$ from the laminar profile occur only in the vicinity of the midplane, where they form chevron-like streaks, see Figs.\ref{pnas_main}(g)-(i). Consequently, the mean velocity profile departs from its laminar counterpart only within a narrow region around the midplane (see Fig.1F of Lellep \emph{et al.} \cite{Lellep2024}). Unlike in Newtonian turbulence, however, the streaks are not sustained by the streamwise vortices that underpin the near-wall cycle \cite{Waleffe1997}: as shown in Figs.\ref{pnas_main}(j)-(l), no significant velocity field develops in the spanwise -- wall-normal plane. Instead, the midplane streaks are accompanied by spanwise-oriented vortices positioned symmetrically above and below the midplane, see Figs.\ref{pnas_main}(g)-(i). These vortices closely resemble those reported in simulations of elasto-inertial turbulence \cite{Samanta2013,Dubief2013,Sid2018,Shekar2019,Shekar2020,Shekar2021}, but whereas in the inertial case they form a part of the wall-mode structures, here they are associated with midplane-localised states. 

\subsection{Characterising elastic turbulence in pressure-driven channel flow}

\begin{figure}[t!]
\centering
\includegraphics[width=1\columnwidth]{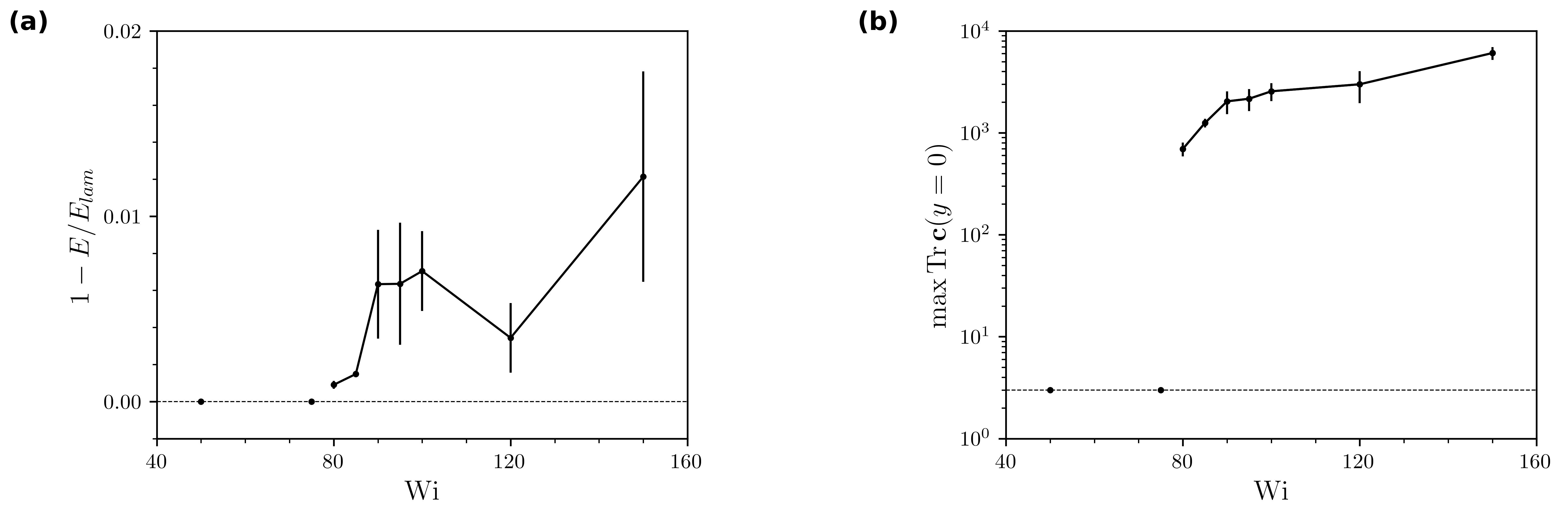}
\caption{Bifurcation diagrams. The finite-amplitude jumps in (a) the reduced kinetic energy and (b) the maximum midplane polymer stretch as functions of $\Wi$. Replotted from the data originally presented in Figs.1D and 1E of Lellep \emph{et al.} \cite{Lellep2024}.}
\label{bif3D}
\end{figure}

The simulations presented above enable us to construct a bifurcation diagram for this flow. To this end, we consider two observables: the total kinetic energy $E(t)$ and the maximum polymer stretch on the midplane, $\max\,\mathrm{Tr}\,{\bm c}(y=0,t)$. Both quantities are monitored throughout the time evolution, and their time averages are evaluated once the system reaches a statistically steady state. Figure \ref{bif3D} shows these averages as functions of the Weissenberg number. In both cases, we observe a sudden jump at $\Wi=80$, marking the onset of elastic turbulence. Close to this threshold, however, relaminarisation events make it difficult to obtain reliable long-time averages due to their stochastic nature.  

A notable difference between the two observables lies in the magnitude of their jumps at the onset. The change in kinetic energy is small, Fig.\ref{bif3D}(a), amounting to only a few percent of its laminar value. Nevertheless, the fluctuations of the reduced kinetic energy remain bounded away from zero, as indicated by the error bars in Fig.\ref{bif3D}(a). Jumps of a comparable magnitude were reported in the microfluidic experiments of Pan \emph{et al.} \cite{Pan2013}, where the onset of elastic turbulence was detected by observing the centerline velocity ﬂuctuations. Such weak signatures in velocity-based observables can easily be mistaken for a linear instability in experiments. In contrast, the midplane polymer stretch exhibits an increase $O(10^3)$ relative to its laminar value, clearly indicating the emergence of a nonlinear state.  

The disparity between the amplitudes of the jumps in $E(t)$ and $\max\,\mathrm{Tr}\,{\bm c}(y=0,t)$ can be understood as follows. For the Reynolds numbers considered here, the Navier-Stokes equation, Eq.\eqref{eq:ns}, effectively reduces to  linear Stokes' equation. As a result, the velocity field is entirely determined by the instantaneous profile of the polymer conformation tensor and does not possess its own time dynamics. In other words, elastic turbulence is driven by the nonlinear dynamics of the polymer stress, while the velocity merely acts as a proxy variable that instantaneously adjusts to the forcing proportional to $\nabla\cdot{\bm c}$. It is important to emphasise that the velocity itself does not directly couple back to the polymer dynamics, except through being trivially involved in the advection term, $\partial {\bm c}/\partial t + {\bm v}\cdot\nabla{\bm c}$. Instead, it is the velocity gradient tensor, $\nabla {\bm v}$, that governs the evolution of $\bm c$. This is quite intuitive from the physical point of view as only the velocity gradient field can stretch an embedded extended object, such as a polymer molecule. Moreover, the dependence of the polymer conformation tensor on $\nabla {\bm v}$ is nonlinear. This is evident even in the laminar state, where the normal stress component scales as $c_{xx}^{\mathrm{(lam)}} \propto \Wi (\partial_y v_x)^2$ in the absence of shear thinning ($\epsilon=0$). Taken together, these ingredients explain why, for sufficiently large $\Wi$, a velocity field with only weak deviations from its laminar profile and moderate gradients can generate large, spatially inhomogeneous stresses, seen in Fig.\ref{bif3D}. This reflects the fundamental fact that the stress (or conformation) tensor is the true dynamical variable, coupled directly to the velocity gradient rather than to the velocity itself.

Statistical features of the velocity and stress fluctuations can be further characterised by their spatial and temporal spectra.  
In Lellep \emph{et al.}~\cite{Lellep2024}, we demonstrated that the one-dimensional spectra of the streamwise velocity fluctuations are broadly consistent with a $k^{-4}$ decay, where $k$ denotes the streamwise wavenumber. In contrast, the spectra of $\mathrm{Tr}\,{\bm c}$ follow a $k^{-2}$ scaling. These power laws are  regarded as hallmarks of chaotic polymeric flows and have been reported in previously~\cite{Steinberg2021,Dubief2013,Garg2021,Steinberg2022}. We also note that the temporal fluctuations of the centreline velocity in our simulations exhibit a power-law decay consistent with $f^{-2}$, where $f$ is the frequency.

Such decay laws differ markedly from the predictions of the Kolmogorov-Richardson theory of Newtonian turbulence~\cite{Batchelor2000}. Beyond the upper bound obtained by Fouxon and Lebedev~\cite{Fouxon2003}, which shows that the velocity spectrum must decay faster than $k^{-3}$, there exists no comprehensive theoretical explanation for the observed exponents. Intriguingly, identical scaling laws have also been reported in active fluids~\cite{Giomi2015,Bardfalvy2019,Alert2020,Weady2022,Negro2025preprint}. Drawing an analogy between the two systems, the observed spectra may be interpreted as spatial fluctuations of the polymer stress around a globally aligned nematic state, comprising highly stretched polymers oriented in the streamwise direction. Since both polymeric and active-fluid systems are governed by equations of motion containing an upper-convected derivative, these exponents may arise as a generic mathematical consequence of this term.

\subsection{Dynamical systems picture of elastic turbulence}
\label{subsec:dynsys}

The majority of the three-dimensional simulations presented above either employed an initial condition constructed from two-dimensional narwhals with a small amount of noise or used a chaotic state obtained at another value of $\Wi$. This choice ensured that the runs quickly reached the chaotic part of the phase space and spent the most computational time sampling the corresponding statistically steady state. However, we have also explicitly demonstrated that a simulation started from the laminar profile and a finite amount of noise in the $c_{xx}$ component of the conformation tensor (generating extensional noise in the midplane) develops into the same statistically steady state. Although the corresponding run spent a long time transitioning from the initial perturbation to elastic turbulence, thus comprising a relatively short turbulent trajectory, it allows us to highlight some important features of the ensuing dynamics. 

\begin{figure}[t!]
\centering
\includegraphics[width=0.45\columnwidth]{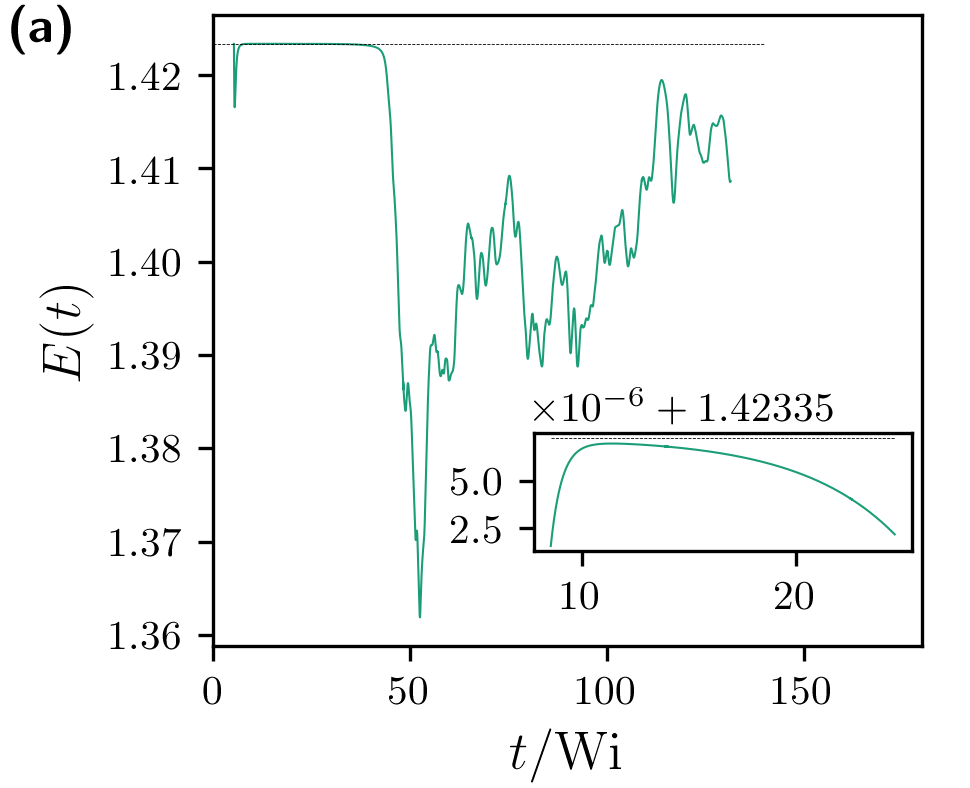}
\includegraphics[width=0.45\columnwidth]{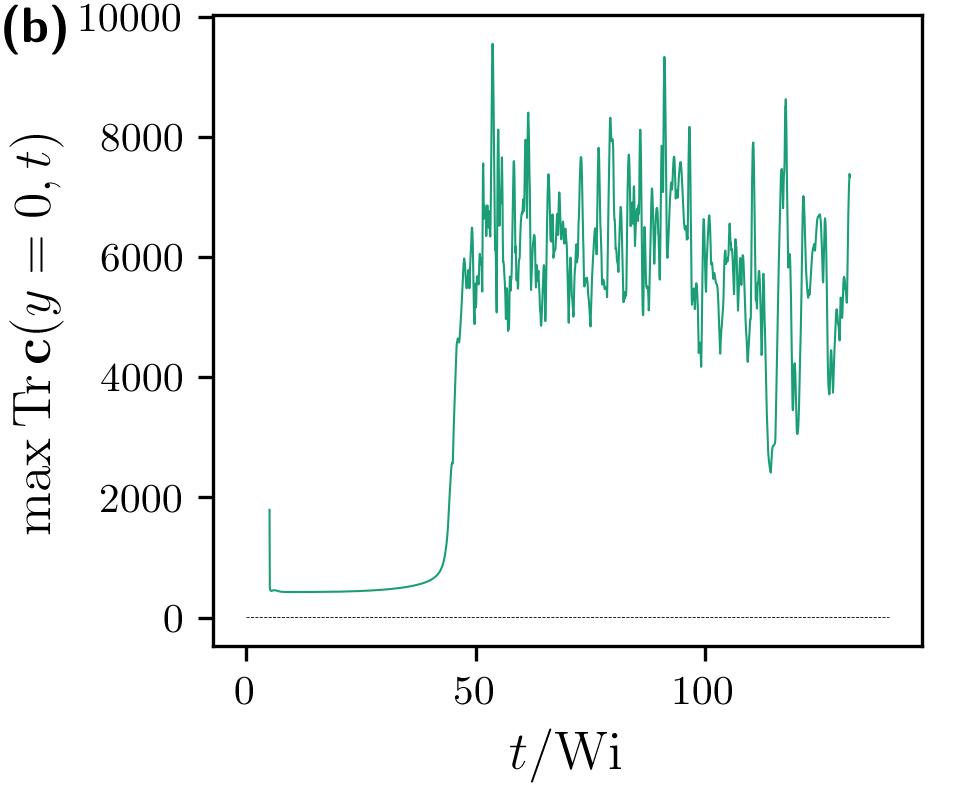}
\caption{Time evolution of a simulation started from a small amount of noise added to the laminar profile at $\Wi=150$: (a) the kinetic energy, (b) the maximum polymer stretch on the midplane. Dashed lines represent the corresponding laminar values. Inset in (a): the short-time evolution of the kinetic energy on its initial approach to the laminar state. 
A part of the data presented in (a) and (b) was previously reported in Figs.1B and S3 of Lellep \emph{et al.}~\cite{Lellep2024}, respectively.}
\label{startupsscalars}
\end{figure}

In Fig.\ref{startupsscalars}, we plot the time evolution of the kinetic energy, $E(t)$, and the maximum polymer stretch on the midplane, $\mathrm{max}\,\mathrm{Tr}\,{\bm c}(y=0,t)$. Initially, the perturbation is decaying with the kinetic energy closely approaching its laminar value. Note, the maximum polymer stretch remains significantly larger than its laminar value throughout this step. In total, the system spends around $40$ polymer relaxation times in the vicinity of the laminar state. After such a long transient, the trajectory finally separates from the vicinity of the laminar attractor and chaotic dynamics ensues. As already discussed above, the total fluctuations of the kinetic energy constitute only a few percent of its laminar value, while the corresponding polymer stress signals are orders of magnitude larger than their laminar counterparts.

\begin{figure}[t!]
\centering
\includegraphics[width=0.6\columnwidth]{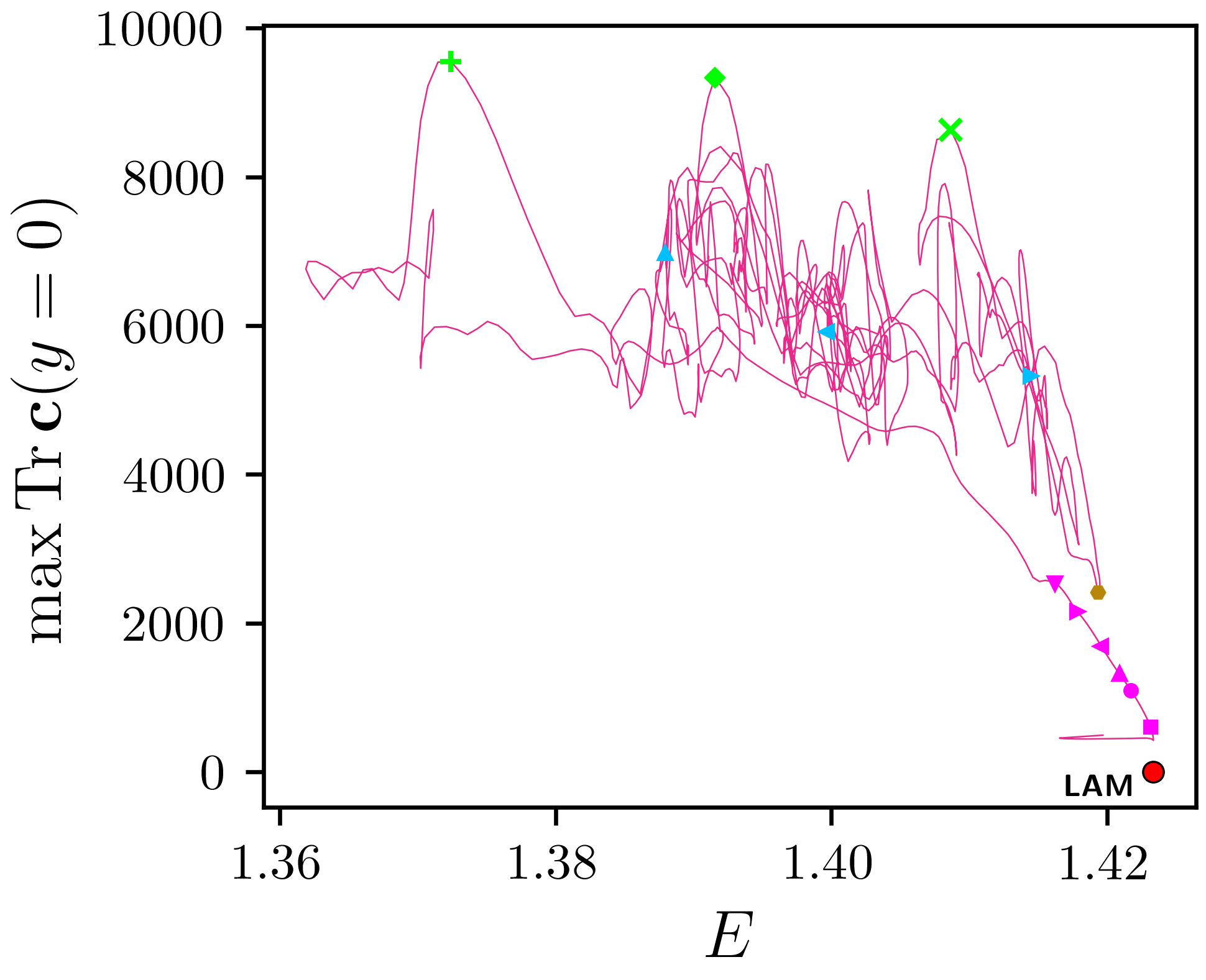}
\caption{A phase portrait of elastic turbulence at $\Wi=150$ constructed from the data in Fig.\ref{startupsscalars}. LAM denotes the position of the laminar state; coloured symbols mark instantaneous states discussed in the text.}
\label{portrait}
\end{figure}

To cast these observations into the dynamical systems language, in Fig.\ref{portrait} we plot a projection of the infinitely dimensional chaotic dynamics of elastic turbulence onto the plane spanned by the instantaneous values of $E$ and $\mathrm{max}\,\mathrm{Tr}\,{\bm c}(y=0)$. First, we note again the region corresponding to the initial evolution of the perturbation as it evolves into the fully developed chaotic state. In that region, we mark several instantaneous states (magenta symbols) that we visualise below. Next, we observe that, away from the laminar state, the trajectory exhibits loops around several points of this two-dimensional space, frequently returning to their vicinity. This observation was originally made in Lellep \emph{et al.}~\cite{Lellep2024} and implies the existence of unstable periodic orbits organising the phase space dynamics of elastic turbulence. Whether this is the case and what these orbits could look like is currently unknown. From these dynamics, we further select several states for closer inspection: the green symbols mark the states corresponding to very high values of the polymer stretch, while the blue ones indicate the parts of the phase space that are visited relatively frequently. Finally, the brown symbol marks a relaminarisation attempt: a close brush-up of the trajectory with the vicinity of the laminar state, see Fig.\ref{startupsscalars} at $t\approx 114$.

\begin{figure}[t!]
\centering
\includegraphics[width=1.0\columnwidth]{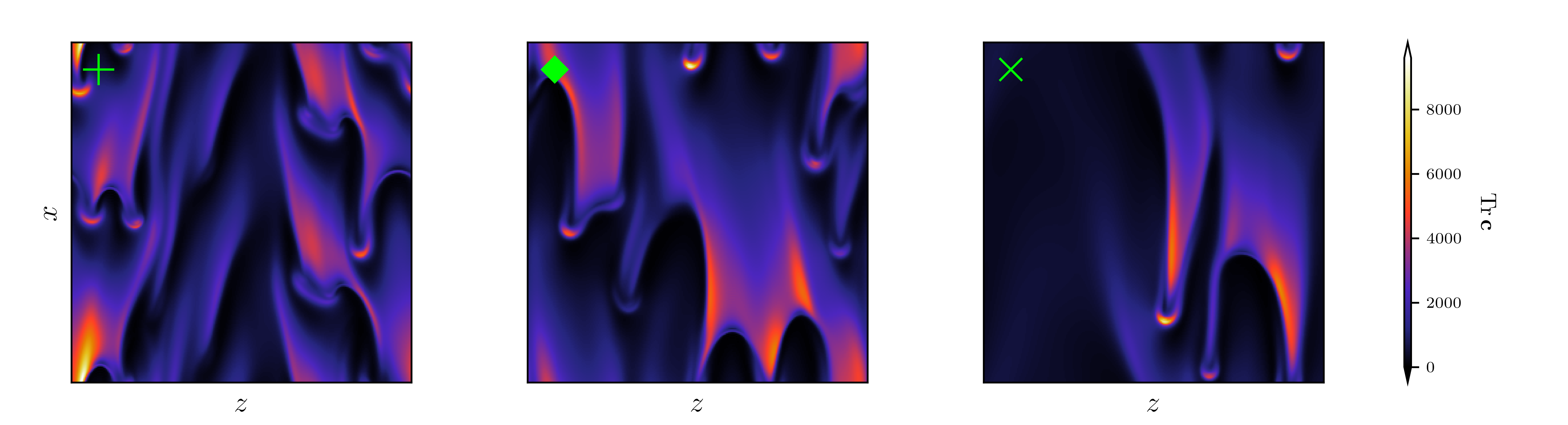}
\includegraphics[width=1.0\columnwidth]{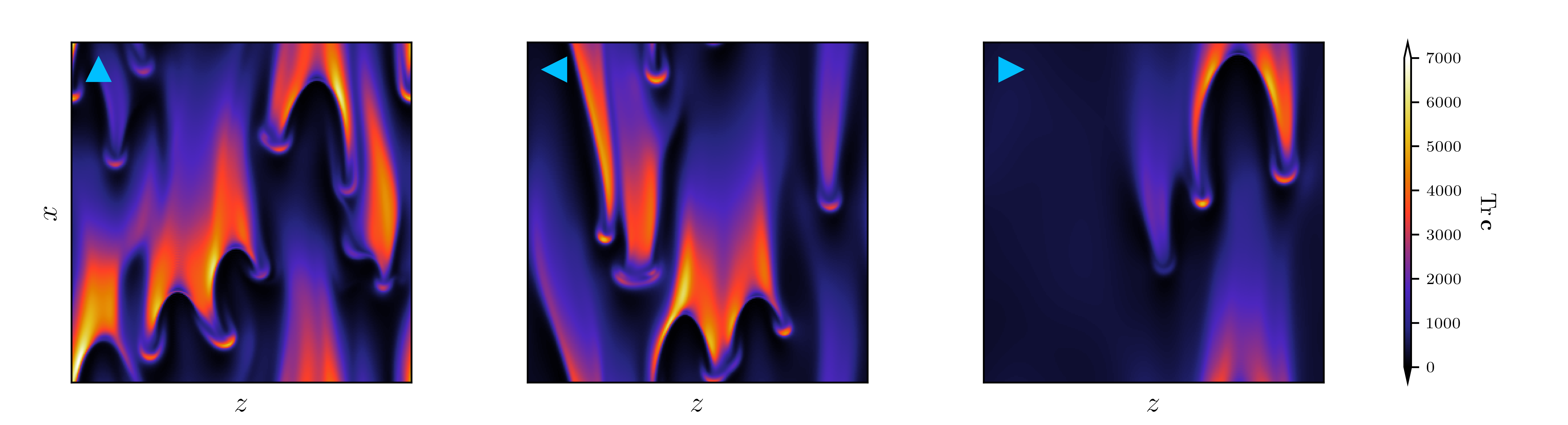}
\caption{Midplane profile of the polymer stretch $\mathrm{Tr}\,{\bm c}$ for the instantaneous states marked in Fig.\ref{portrait}. The mean flow is from bottom to the top along the $x$-direction. }
\label{topsmids}
\end{figure}

Visualisations of the `green' and `blue' states, Fig.\ref{topsmids} reveal a complicated spatio-temporal intermittency with the number of spatially localised states, three-dimensional narwhals, fluctuating in time. 
The low values of the kinetic energy typically indicate the presence of a large number of such structures in the domain (left and middle rows in Fig.\ref{topsmids}), while higher values of the kinetic energy (right row) invariably corresponds to a simplification in the dynamics and a low number of structures in the domain. This observation is readily rationalised by noting that the extensional nature of the narwhal states discussed in Section \ref{subsec:explanation} requires the flow to stretch polymers along the midplane thus draining some kinetic energy from the system; note that $E(t)$ is therefore bound from above by its laminar value.
Interestingly, the outlier states corresponding to the highest values of the polymer stretch (green symbols) are structurally similar to the more usual, `blue' states, with only a small part of the domain exhibiting very high values of $\mathrm{Tr}\,{\bm c}$. Such intermittency can also be directly observed in the Supplemental Movies S2 and S3 of Lellep \emph{et al.}~\cite{Lellep2024}.

\begin{figure}[t!]
\centering
\includegraphics[width=1.0\columnwidth]{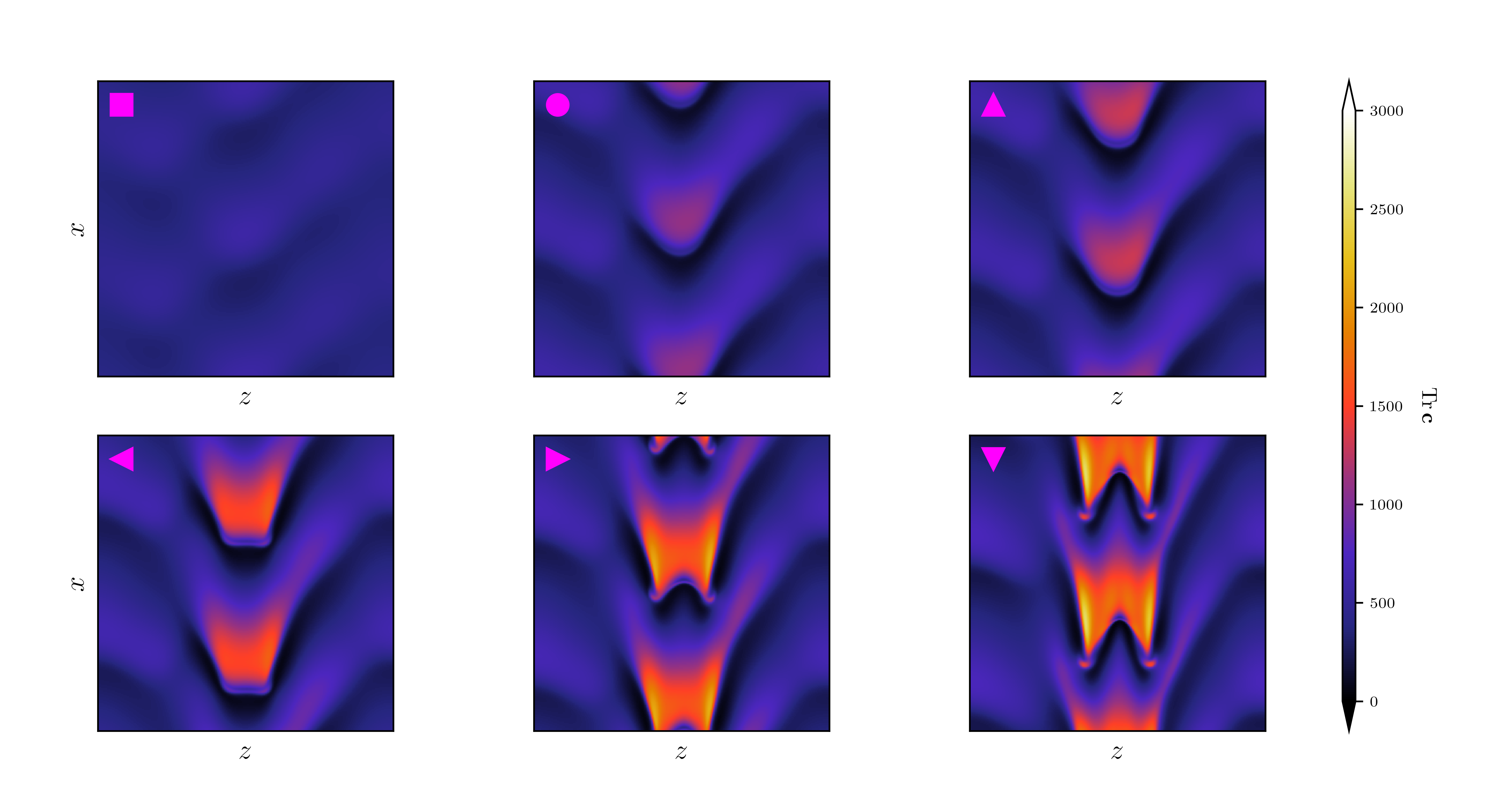}
\caption{Midplane profile of the polymer stretch $\mathrm{Tr}\,{\bm c}$ for the early time (magenta) states in Fig.\ref{portrait}. The mean flow is from bottom to the top along the $x$-direction. }
\label{startups}
\end{figure}

Visualising the early time development of the perturbation in Fig.\ref{startups} allows us to understand the structure of the edge state separating the chaotic dynamics of elastic turbulence from the basin of attraction of the laminar flow. The original perturbation comprising small-scale noise everywhere on the domain quickly decays (not shown), with only a few, relatively large-scale regions of high stress remaining the first snapshot in Fig.\ref{startups}. As time progresses, a particular spanwise position is selected in the midplane through spontaneous symmetry-breaking and a train of blob-like structures emerges. These structures become progressively blunt at their trailing edge, eventually developing a parabolic region of high polymer stretch enveloping the region of almost no stretch; this arrangement is a typical signature of a midplane cut through a three-dimensional structure in Fig.\ref{3dprinted}. At later times, the symmetry of the train is broken yet again (not shown) and a fully chaotic dynamics takes over the entire domain.

\begin{figure}[t!]
\centering
\includegraphics[width=0.33\columnwidth]{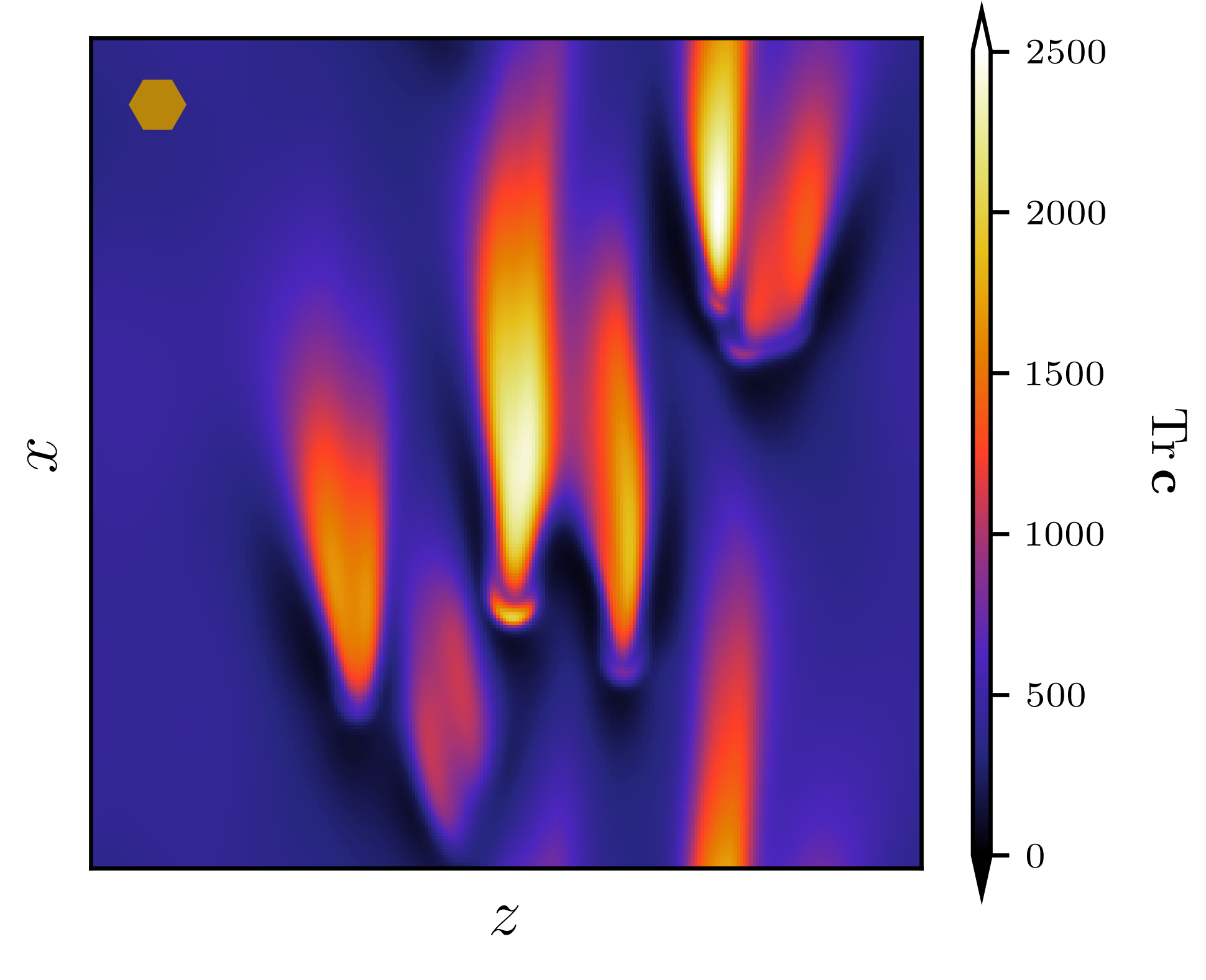}
\caption{Midplane profile of the polymer stretch $\mathrm{Tr}\,{\bm c}$ corresponding to the relaminarisation attempt (brown marker) in Fig.\ref{portrait}. The mean flow is from bottom to the top along the $x$-direction. }
\label{edge}
\end{figure}

Remarkably, the structures shown in Fig.\ref{startups} share structural similarities with the state visualised in Fig.\ref{edge} observed at a very late time, $t\approx 114$. We interpret the latter as a relaminarisation attempt, characterised by the low values of the polymer stretch, the values of the kinetic energy approaching its laminar value, and the number of localised states decreasing at this point. These similarities indicate that both instances correspond to the trajectory approaching the edge state from different directions (the laminar and the turbulent sides in Figs.\ref{startups} and \ref{edge}, respectively). Additionally, since the publication of Lellep \emph{et al.}~\cite{Lellep2024}, we have collected a significant number of observations of various turbulent trajectories occasionally approaching the vicinity of a state strongly resembling that in Figs.\ref{startups} and \ref{edge} at various values of $\Wi$ and $\Re$, and we are confident that they correspond to the vicinity of the edge state of elastic turbulence. Isolating such a state and understanding the mechanism that sustains it, similar to Section \ref{subsec:explanation}, will be addressed in our future work.

\subsection{A box full of narwhals}
\label{sect:longbox}

\begin{figure}[ht!]
\centering
\includegraphics[width=1\columnwidth]{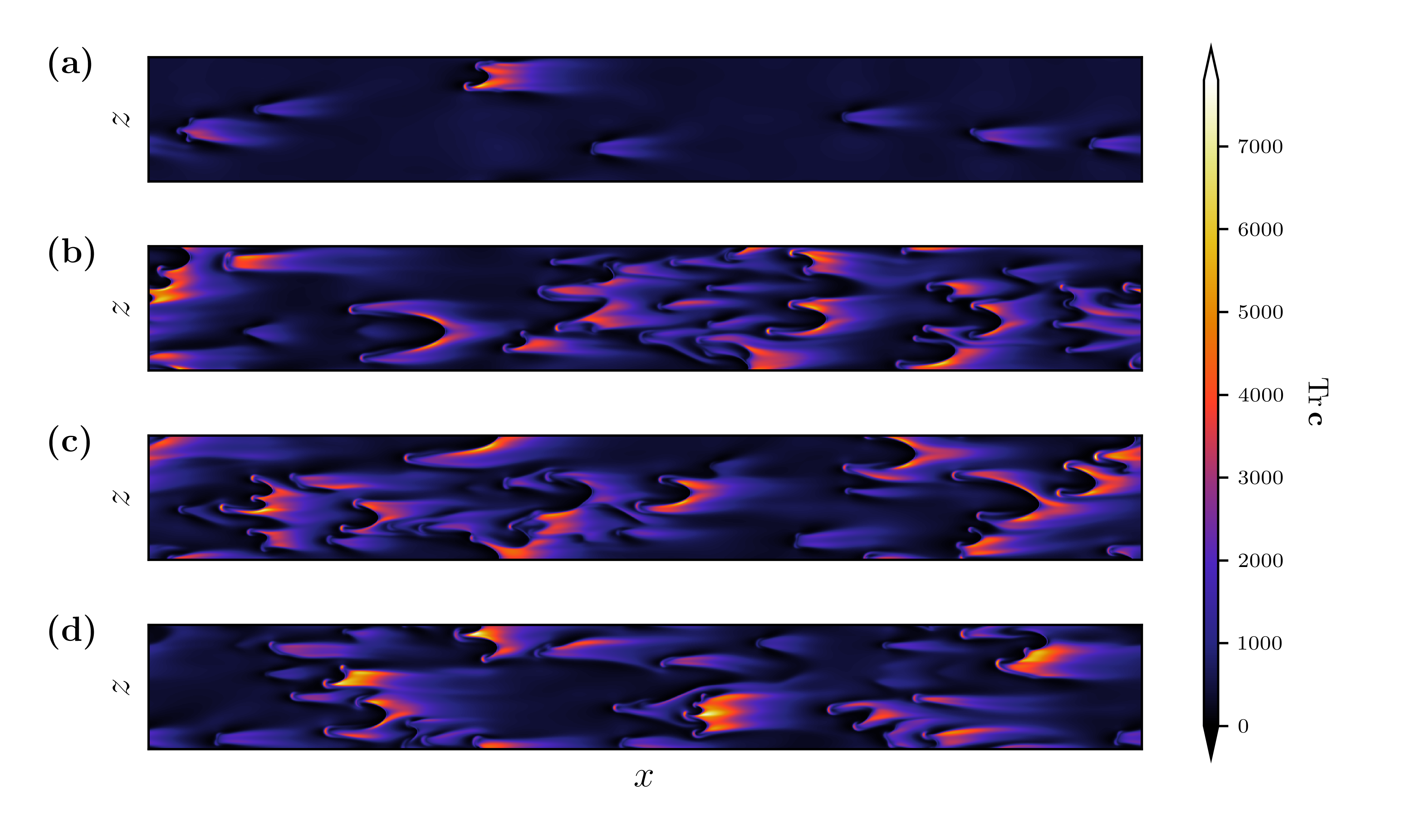}
\caption{Midplane profiles of the polymer stretch in a long box with $L_x=80$ at $\Wi=150$. Panels (a)-(d) represent different instances along the turbulent trajectory, with time increasing from (a) to (d).}
\label{LongBlessings}
\end{figure}

The results discussed so far were obtained in relatively small computational domains with $L_x=L_z=10$. As seen in the case of two-dimensional narwhals, structures sustained by extensional stresses along the midplane possess a characteristic length scale and constraining them to relatively small boxes may therefore strongly affect their dynamics. To address this, here we examine the influence of the domain size on the phenomenology reported above. Specifically, we present a preliminary simulation performed in an extended box with $L_x=80$ and $L_z=10$ at $\Wi=150$. Owing to the prohibitive numerical cost, this run was continued for only about $40$ relaxation times (measured as $t/\Wi$) after reaching the statistically steady state. While shorter than the intervals used in Fig.~\ref{startupsscalars}, this duration remains sufficient for a meaningful comparison between the two cases.

In Fig.~\ref{LongBlessings}, we show several instantaneous snapshots from this simulation. Panel~(a) corresponds to the initial transient, while panels~(b)-(d) were recorded in the statistically steady state. Firstly, we note that the structures observed in Fig.~\ref{LongBlessings} are comparable in size to those in Fig.~\ref{topsmids}, indicating that a domain with $L_x=10$ is already sufficient to capture their dynamics. Secondly, as in the small box, both the early time transient, Fig.~\ref{LongBlessings}(a), and the instantaneous profiles in the steady state exhibit excursions towards the edge state discussed in Section~\ref{subsec:dynsys}. Unlike the small-box dynamics, however, these excursions now appear as spatially localised events, while the rest of the domain remains turbulent. Thirdly, the extended domain reveals clusters of coherent states separated by nearly laminar regions along the streamwise direction. Such spatially intermittent dynamics are a hallmark of Newtonian parallel shear flows \cite{Avila2011,Barkley2016,Avila2023}, and their significance in the present context will be revisited in Section~\ref{sec:discussion}. Finally, Figs.~\ref{LongBlessings}(b)-(d) show that the localised states undergo splitting, reminiscent of Newtonian turbulence \cite{Shimizu2014,Barkley2016,Avila2023} and of the recent observations of Shnapp and Steinberg \cite{Shnapp2022b} in polymeric channel flows. Interestingly, the present results suggest that such splitting events may also occur along the spanwise direction. The precise nature and mechanism of these events will be investigated in future work.

\section{Discussion}
\label{sec:discussion}

The results presented in this paper establish localised coherent states as the building blocks of elastic turbulence in parallel shear flows of model polymeric fluids. Since the flow is linearly stable across the wide parameter range considered here, these states emerge through a subcritical bifurcation from infinity. In two-dimensional flows, the nonlinear states, the narwhals, are travelling-wave solutions organised around a solitary pair of stagnation points along the centreline and accompanied by a pair of vortices near the tusk--body junction. The narwhals are not governed by the hoop-stress instability and their appearance is not captured by the Pakdel-McKinley criterion. Although there is currently no established framework to understand them, our results suggest that they are sustained by a novel mechanism based on extensional stresses. In long simulation boxes, they appear to localise along the flow direction, indicating the existence of a characteristic streamwise lengthscale.  

When embedded in three-dimensional simulation boxes, narwhals lose stability and transition to a chaotic flow state characterised by coherent, spot-like structures organised around the channel midplane. These structures have only a mild effect on velocity-based observables, with the midplane velocity fluctuating only a few percent around its laminar value, but they strongly impact polymer stress fluctuations, consistent with the stress being the driving dynamical quantity of elastic turbulence. Strikingly, instantaneous stress profiles in the streamwise--wall-normal plane often reveal local traces of narwhal-like structures. This suggests that, although linearly unstable, two-dimensional narwhals continue to organise the dynamics of three-dimensional flows. Above the instability threshold we observed sudden relaminarisation events, while at sufficiently high $\Wi$ we reported \emph{blessings} of multiple spot-like structures, with their number fluctuating significantly in time.  

These results represent the first computational observations of elastic turbulence in parallel shear flows. Achieving them is a milestone in a field long plagued by the High-Weissenberg Number Problem -- a suite of numerical instabilities that made simulations at $\Wi>1$ notoriously difficult. The advances reported here were enabled by the discovery of the centreline instability by Shankar and colleagues \cite{Garg2018,Chaudhary2019,Khalid2021a,Chaudhary2021}, the pioneering two-dimensional computational work of Berti \emph{et al.}~\cite{Berti2008,Berti2010} and Page \emph{et al.}~\cite{Page2020}, and the power of modern supercomputers that allow simulations at very high spatial resolution (see \cite{Lin2025} for a recent example).

The field of elastic instabilities and turbulence in parallel shear flows is still in its infancy. Its current status is comparable to that of Newtonian turbulence at the dawn of the first direct numerical simulation. Unsurprisingly, there are more open questions than answers. Below we highlight several research directions that, in our view, are especially important, many of them inspired by recent progress in Newtonian shear flows.  

\begin{itemize}
\item \emph{Self-sustaining mechanism.} What is the mechanism that sustains three-dimensional narwhals? We expect it to be a version of the extensional-stress argument presented in Section~\ref{subsec:explanation}, but this has not yet been developed.  

\item \emph{Lengthscales of elastic turbulence.} What sets the characteristic lengthscales of elastic turbulence? Section~\ref{subsec:explanation} suggests that the key role is played by shear-thinning effects \cite{Becherer2008}.  

\item \emph{Minimal flow unit.} Is there a minimal flow unit of elastic turbulence, analogous to the Newtonian \cite{Jimenez1991} and two-dimensional viscoelastic \cite{Zhang2024} cases? The extensional nature of narwhals (Section~\ref{subsec:length}) suggests that their individual length increases with $\Wi$, implying that the minimal unit could paradoxically grow larger as turbulence develops.  

\item \emph{Unstable periodic orbits.} The phase portrait in Fig.~\ref{portrait} strongly suggests the presence of unstable periodic orbits. Can they be isolated and characterised?  

\item \emph{What is the edge state of elastic turbulence?  } 

\item \emph{Directed percolation transition?} We observed localised states that can relaminarise, merge, and split. Similar splitting events were reported experimentally in transitional elastic flows \cite{Shnapp2022b}. In Newtonian shear flows, such processes underpin a transition to space-filling turbulence that belongs to the directed percolation universality class \cite{Hinrichsen2000,Barkley2016,Lemoult2016,Sano2016}. Is a similar scenario possible in the purely elastic case?  

\item \emph{Space-filling elastic turbulence?} 
What is the structure of the stress field in elastic turbulence at large $\Wi$? We do not have a clear prediction based on our simulations. On the one hand, 
the analogy with the Newtonian case implies that splitting should eventually dominate over relaminarisation, producing a space-filling state. On the other hand, however, the narwhal solutions are supported by a solitary pair of stagnation points and such a structure is not easily merged into a continuum.
Perhaps high-$\Wi$ elastic turbulence instead resembles the blessing in Fig.~\ref{LongBlessings}, with the spatial density of spot-like structures increasing with $\Wi$.  

\item \emph{Diffusive instability.}  Recent work by Kerswell and colleagues demonstrates that the presence of stress diffusivity can trigger a novel linear \emph{diffusive instability} \cite{Beneitez2023,Couchman2024,Lewy2024,Beneitez2025}. Although originally identified in models with explicit stress diffusion, it has since been extended to numerical schemes without such terms. The diffusive instability sets in on a scale comparable to a distance diffused by a single polymer molecule in one relaxation time; it is therefore microscopic in nature and we view it as unphysical. Unfortunately, its presence makes some parts of the parameter space inaccessible to direct numerical simulations and there is a pressing need to develop amended polymeric equations of motion that suppress this instability.

 \item \emph{Experimental verification.}  The field is in dire need of quantitative experimental verification. The results presented here indicate that, close to the onset of elastic turbulence, one can expect only modest levels of fluctuations in velocity-based observables. Although such weak signals were successfully measured by Pan \emph{et al.}~\cite{Pan2013}, their elusiveness represents a major challenge. The situation might improve at much larger values of $\Wi$, or at higher $\Re$. Recent work on elasto-inertial turbulence indicates that such flows are dominated by wall-mode structures \cite{Samanta2013,Dubief2013,Sid2018,Shekar2019,Shekar2020,Shekar2021} with a significantly higher level of fluctuations in the velocity-based observables \cite{Choueiri2018,Lopez2019,Zhu2020,Choueirie2021,Rota2024}. One might therefore expect a crossover from the centre-mode dominated purely elastic dynamics to wall-mode dominated flows at higher $\Re$ \cite{Rota2024}. Around this transition, both regimes may yield stronger signals, aiding experimental identification. In parallel, it is highly desirable to develop the ability to directly measure polymer stress, resolved in time and space.  
 
\item \emph{Universality of narwhal states.} In addition to pressure-driven channel flow discussed here, narwhal/arrowhead states have now been reported in Kolmogorov flow \cite{Berti2008,Berti2010,Nichols2025,Lewy2025}, in a flow past an array of cylinders \cite{Zhu2024}, and in flows of highly entangled polymers and wormlike micelles\cite{Lewy2025preprint}, while we recently obtained first results on narwhal-like states in pipe flows. 
This suggests that such structures may be universal. Identifying the general conditions for their existence is an important open question.  

\end{itemize}

We expect these and related questions to shape our research programme on elastic turbulence for years to come.  

\begin{acknowledgments}
The authors would like to thank Paulo Arratia, Keaton Burns, Bj\"{o}rn Hof, Sandra Lerouge, Anke Lindner, Davide Marenduzzo, Jeff Oishi, Rob Poole, Wim van Saarloos, Becca Thomases, Geoff Vasil, and Christian Wagner  
for their advice and continuing support. This work used the ARCHER2 UK National Supercomputing Service (https://www.archer2.ac.uk). Support from the UK Turbulence Consortium (EPSRC grants EP/R029326/1 and EP/X035484/1) and ARCHER2 team are gratefully acknowledged. Martin Lellep was supported by the German Academic Scholarship Foundation (Studienstiftung des deutschen Volkes). We acknowledge financial support from the Priority Programme SPP 1881 ``Turbulent Superstructures" of the Deutsche Forschungsgemeinschaft (DFG) under grant Li3694/1. 

For the purpose of open access, the authors have applied a Creative Commons Attribution (CC BY) licence to any Author Accepted Manuscript version arising from this submission.
\end{acknowledgments}

\appendix

\section{Dimensionless units}
\label{app:units}

The governing equations are rendered dimensionless by using the following characteristic scales: all lengths are scaled with the channel half-width $d$, velocities with $U_0$, time with $d/U_0$, and pressure with $(\mu_s+\mu_p)U_0/d$. Here, $\mu_s$ and $\mu_p$ are the solvent and polymeric contributions to the viscosity, respectively, while the velocity scale $U_0$ corresponds to the  laminar centreline velocity of a Newtonian fluid with the viscosity $\eta_s+\eta_p$ at the same value of the applied pressure gradient.

The flow is controlled by several dimensionless quantities. The strength of fluid elasticity is described by the Weissenberg number, $\Wi=\lambda\,U_0 / d$, the relevance of inertia - by the Reynolds number, $\Re=\rho\,U_0 d/(\mu_s+\mu_p)$, the relative contribution of the polymers to the fluid viscosity - by the ratio $\beta = \eta_s/(\eta_s + \eta_p)$, the degree of shear-thinning is set by $\epsilon$, while the stress diffusivity, $\kappa = D / d\,U_0$, compares the typical distance diffused by a polymer molecule in one time unit with the channel half-width. Here, $\rho$ is the density of the fluid, $\lambda$ is its Maxwell relaxation time, and $D$ is the diffusion coefficient of a polymer molecule. 
Throughout this paper, time intervals are reported in terms of $t/\Wi$, which in our dimensionless units corresponds to measuring physical time in terms of the Maxwell relaxation time $\lambda$.

\bibliography{et}

\end{document}